\documentclass[5p,twocolumn,10pt,times]{elsarticle}
\usepackage{amsmath}
\usepackage{hyperref}
\usepackage{algorithm}
\usepackage{algpseudocode}
\usepackage{multirow}
\usepackage{microtype}
\usepackage{amsfonts}
\usepackage{graphicx}
\usepackage{booktabs}
\usepackage{lineno}
\usepackage[final]{pdfpages}
\usepackage{xcolor}
\begin{document}
\baselineskip11pt
\setpagewiselinenumbers
\begin{frontmatter}

\title{Winding clearness for differentiable point cloud optimization}

\author[1,4]{Dong Xiao}
\ead{xiaodong@ustc.edu.cn}

\author[2,3]{Yueji Ma}
\ead{myj21@mails.tsinghua.edu.cn}

\author[3,5]{Zuoqiang Shi}
\ead{zqshi@tsinghua.edu.cn}

\author[6]{Shiqing Xin}
\ead{xinshiqing@sdu.edu.cn}

\author[7]{Wenping Wang}
\ead{wenping@tamu.edu}

\author[8]{Bailin Deng}
\ead{dengb3@cardiff.ac.uk}

\author[1]{Bin Wang\corref{cor}}
\ead{wangbins@tsinghua.edu.cn}

\cortext[cor]{Corresponding Author}

\address[1]{School of Software, Tsinghua University, China}
\address[2]{Department of Mathematical Sciences, Tsinghua University, China}
\address[3]{Yau Mathematical Sciences Center, Tsinghua University, China}
\address[4]{School of Mathematical Sciences, University of Science and Technology of China, China}
\address[5]{Yanqi Lake Beijing Institute of Mathematical Sciences and Applications, China}
\address[6]{School of Computer Science and Technology, Shandong University, China}
\address[7]{Department of Computer Science and Engineering, Texas A\&M University, USA}
\address[8]{School of Computer Science and Informatics, Cardiff University, Wales, UK}

\begin{abstract} 
We propose to explore the properties of raw point clouds through the \emph{winding clearness}, a concept we first introduce for measuring the clarity of the interior/exterior relationships represented by the winding number field of the point cloud. In geometric modeling, the winding number is a powerful tool for distinguishing the interior and exterior of a given surface $\partial \Omega$, and it has been previously used for point normal orientation and surface reconstruction. In this work, we introduce a novel approach to evaluate and optimize the quality of point clouds based on the winding clearness. We observe that point clouds with less noise generally exhibit better winding clearness. Accordingly, we propose an objective function that quantifies the error in winding clearness, solely utilizing the coordinates of the point clouds. Moreover, we demonstrate that the winding clearness error is differentiable and can serve as a loss function in point cloud processing. We present this observation from two aspects: 1) We update the coordinates of the points by back-propagating the loss function for individual point clouds, resulting in an overall improvement without involving a neural network. 2) We incorporate winding clearness as a geometric constraint in the diffusion-based 3D generative model and update the network parameters to generate point clouds with less noise. Experimental results demonstrate the effectiveness of optimizing the winding clearness in enhancing the point cloud quality. Notably, our method exhibits superior performance in handling noisy point clouds with thin structures, highlighting the benefits of the global perspective enabled by the winding number. The source code is available at~\url{https://github.com/Submanifold/WindingClearness}.

\end{abstract}


\end{frontmatter}


\section{Introduction}\label{sec:introduction}

The winding number serves as a valuable tool for determining the inside/outside position of a query point $\textbf{q} \in \mathbb{R}^3$ relative to a given closed surface $\partial \Omega$. This tool has found widespread application in various geometric tasks~\cite{2013Winding, 2022Medial, 2022RDT, 2023WindingDiscrete}. In recent years, there has been a growing research effort into its applicability for point cloud processing. Existing studies mainly focus on leveraging the winding number theory for normal orientation and surface reconstruction from point clouds~\cite{2019GR, 2022PGR, 2021dipole, 2023GCNO, 2024WNNC}.

In this work, we investigate a new application of the winding number theory in evaluating and enhancing the quality of point clouds. Point clouds frequently suffer from issues such as noise and misalignment, as reported in~\cite{2017ReconSurvey}. These issues can affect the quality and accuracy of their geometric representation. The need to enhance the quality of point clouds is common in various scenarios. For example, in the field of 3D deep learning, the 3D diffusion model has been widely used in point cloud generation tasks~\cite{2021DPM, 2021PVD, 2022LION, 2023FastDiffusion}. However, the generated point clouds usually have relatively low quality for subsequent use. The datasets utilized for 3D generation typically feature thin structures. Many traditional denoising methods utilize the nearest neighbors of a point to model its local neighborhood on the underlying surface and analyze its geometric properties. However, such a local approach may fail on thin structures, as the nearest neighbors may include points from the other side of the surface that do not belong to the local neighborhood. In this work, we leverage the global awareness of the winding number to offer a fresh perspective on these problems.

Our main observation is that a high-quality point cloud, for instance, with a low noise level, tends to exhibit improved clarity in the internal/external relationship determined by the winding number field, demonstrating better \emph{winding clearness}. Specifically, the implicit function values of the sample points approaches the iso-value 1/2. Meanwhile, the value in the bounding box, slightly larger than the shape, approaches zero. We define the difference between the actual values and the expected values of the winding number field at these specific points as the \emph{winding clearness error}. The point clouds with less noise typically exhibit lower winding clearness errors. 

\begin{figure*}[htb]
  \centering
  \includegraphics[width=1.0\linewidth]{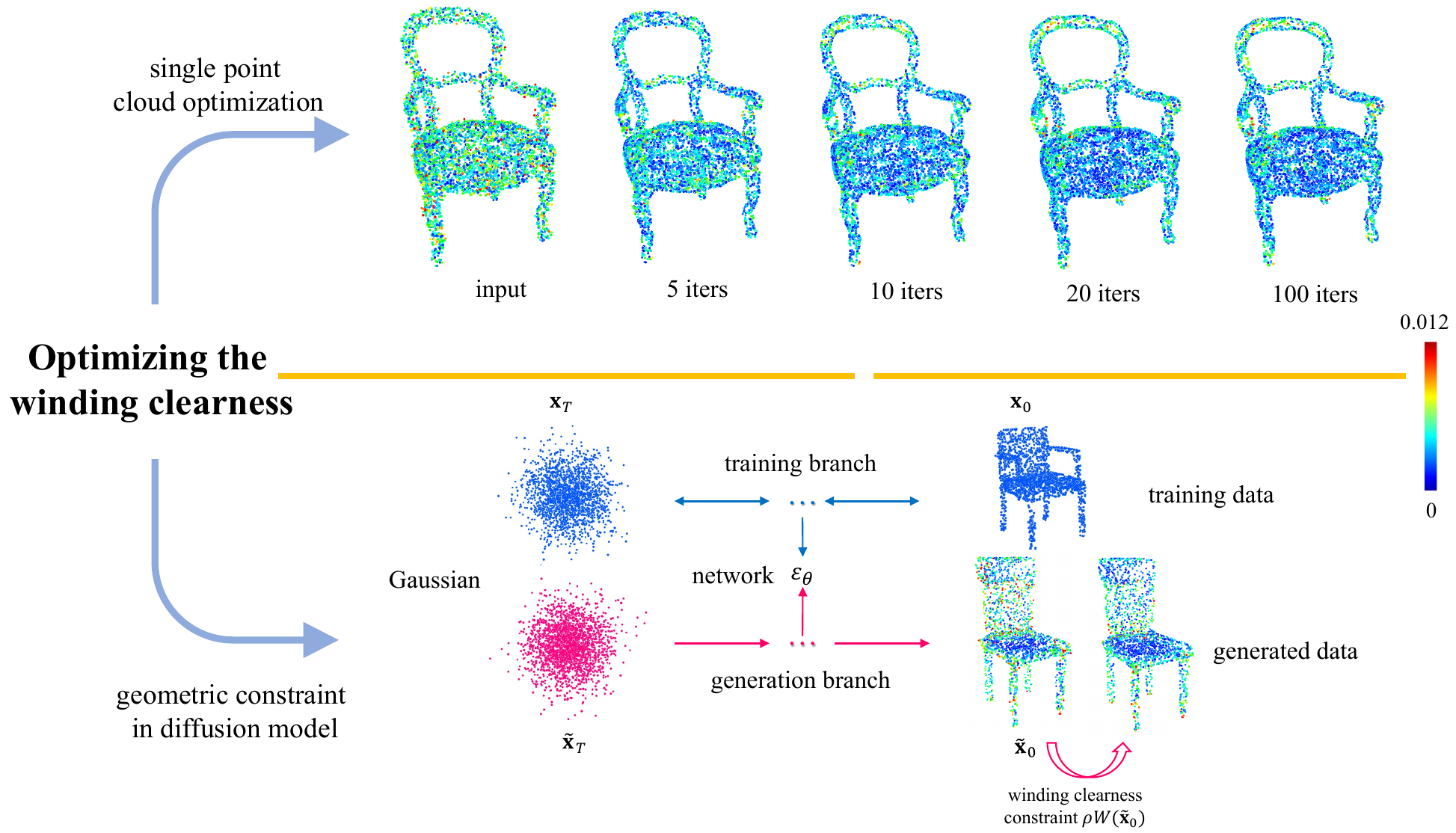}
  \caption{\label{fig:overview} The methodology overview of our work. We present two approaches to optimize the winding clearness of point clouds. 1) Direct optimization of a single point cloud in a non-data-driven style. 2) Incorporating winding clearness as a geometric constraint in diffusion-based 3D point cloud generation through a joint training strategy. 
           }
\end{figure*}

Additionally, as the winding clearness error is a differentiable function that solely depends on the point coordinates, we can treat it as a loss function for point cloud optimization and integrate it into point processing pipelines that require differentiability. We demonstrate this observation from two perspectives: 1) We directly back-propagate the loss function to optimize point positions without relying on a neural network. 2) We incorporate winding clearness as a geometric constraint in the diffusion-based 3D point cloud generation framework, enabling fine-tuning of the generative model and improving the quality of the resulting point clouds. This approach successfully bridges the integrated implicit field with the point cloud topology. The methodology overview of our work is depicted in Figure~\ref{fig:overview}, with further details provided in Section~\ref{sec:Optimization}.

In the experiments, we demonstrate that improving the winding clearness through both two approaches enhances the quality of the point cloud. We conduct experiments on various datasets and compare our methods with related approaches. The results indicate that our method exhibits competitive performance on traditional geometric models. Moreover, our method demonstrates superior performance in handling noisy point clouds with thin structures, surpassing traditional methods that rely on local neighborhood information. This finding highlights the effectiveness of the global perspective brought by the winding number.

\section{Related work}\label{sec:Related_Work}

In this section, we provide a review of relevant studies into three aspects: the utilization of the winding number in point cloud processing, the management of low quality point clouds, and the generation of 3D point clouds utilizing the diffusion-based technique.

\subsection{Winding number for geometry processing of point clouds}
The winding number has proven to be a valuable tool in 3D geometry processing for robust inside/outside judgement as demonstrated in previous studies~\cite{2013Winding, 2018FastWinding}. This concept can also be understood by the Gauss formula in potential energy theory~\cite{1995PDEbook, 2019GR, 2022PGR} or the electronic dipole field~\cite{2021dipole}, and can also be extended to discrete surfaces with non-trivial topology~\cite{2023WindingDiscrete}. While the winding number is commonly utilized in mesh processing tasks, such as voxelization~\cite{2018FastWinding}, boolean operations~\cite{2016Solid}, facet orientation~\cite{2014FacetsOrientation} and medial axis transform~\cite{2022Medial}, our focus is on exploring its application to point processing. 

Approaches such as fast winding~\cite{2018FastWinding} and GR~\cite{2019GR} leverage the winding number for various point cloud tasks, including surface reconstruction from oriented point sets. They also utilize the fast multipole method (FMM)~\cite{1987FMM} to improve the computational efficiency. Chen et al.~\cite{2024FastDipole} expand this technique to neural rendering and achieves high-quality 3D reconstruction from multi-view images. Some other research focuses on normal orientation and surface reconstruction from unoriented point clouds, relying on the global awareness achieved by the winding number, such as PGR~\cite{2022PGR}, GCNO~\cite{2023GCNO} and WNNC~\cite{2024WNNC}. Specifically, PGR introduces an innovative approach that eliminates the reliance on surface normals and achieves unoriented surface reconstruction by treating the surfels (which represent the multiplications of point normals and surface elements) as unknowns and solving a linear system to determine the magnitude and direction of the surfels. GCNO addresses globally consistent normal orientation by regularizing the winding number field with a series of energy functions. WNNC presents an iterative method for winding number normal consistency and achieves fast and global normal orientation by leveraging the FMM technique. However, the point positions in the aforementioned methods remain static. In contrast, our approach presents a novel perspective by enhancing the quality of point clouds via winding clearness optimization.

\subsection{Management of low quality point clouds}
Point clouds are commonly used data structures for representing geometric shapes. However, they frequently suffer from artifacts such as noise and misalignment~\cite{2017ReconSurvey}. Denoising is a widely explored approach to address this issue and improve the quality of point clouds. This problem has been extensively studied for several decades, as reviewed in~\cite{2017ReviewFilter} and~\cite{2022PointCloudReview}. 

Several traditional denoising methods primarily concentrate on local filtering and rely on the nearest neighbor computation. Digne et al.~\cite{2017Bilateral} and Zhang et al.~\cite{2019BilateralPCA} employ the bilateral filter for 3D point cloud denoising. Cazals et al.~\cite{2005Jets} approximates local properties through the osculating jets and projects the noisy points onto the approximated local shape. LOP~\cite{2007LOP} and WLOP~\cite{2009WLOP} propose a local projection strategy and iteratively refine a point cloud subset, serving as a denoised representation of the original point set. However, these methods often struggle with thin structures due to their reliance on local neighbors. Other denoising methods are based on different mechanisms, such as moving least squares (MLS) surfaces~\cite{2003PointSetSurface, 2005MovingSharp, 2009KernelRegre, 2007APSS}, $L_{1}$ and $L_{0}$ regularization~\cite{2010l1sparse, 2015l0sparse}, Taubin smoothing~\cite{2022Elliptic} and low rank optimization~\cite{2022LowRank}. Nevertheless, the effectiveness of previous methods in resolving this problem is still limited. Some techniques employ deep neural networks for point cleaning and achieve promising results~\cite{2020Pointcleannet, 2021Pointfilter, 2022MODNet, 2023PCDNF, 2023IterativePFN, 2024SharpNeural}. However, they usually require a large amount of training data. Other methods primarily concentrate on preserving sharp or edge features, exemplified by EAR~\cite{2013EdgeAware} and RFEPS~\cite{2022RFEPS}. However, these methods do not notably change the denoising process in non-feature areas. Additionally, the input point sets in our method are unoriented, which sets our method apart from mesh denoising methods (e.g., ~\cite{2018MeshDenoising}). Those methods have an inherent orientation and can utilize normal information.

\subsection{Diffusion-based 3D generation of point clouds}
In recent years, with the growing popularity of 3D generative models, especially the Denoising Diffusion Probabilistic Model (DDPM)~\cite{2020DDPM}, significant advancements have emerged in diffusion-based point cloud generation. The traditional DDPM entails learning a probabilistic denoising procedure, including both a reverse process and a generative process. DPM~\cite{2021DPM} models the reverse process as a Markov chain operating on the latent space of shapes. PVD~\cite{2021PVD} utilizes a point-based network for generating point clouds from an initial Gaussian prior and intermediate noises that matches the dimension of the output. LION~\cite{2022LION}, proposed by NVIDIA, introduces a hierarchical generation process that combines the global shape latent with the point-structured latent. PSF~\cite{2023FastDiffusion} proposes to generate point clouds via the transport flow, and distills the flow for single-step generation. However, prior research in this domain predominantly focuses on ensuring the similarity between the generated point sets and the original dataset, with limited specific consideration for the quality of the generated point clouds. To fill this gap, we propose the incorporation of a differentiable point cloud optimization technique for this problem. Some alternative approaches can directly generate triangle meshes from the diffusion process~\cite{2023DiffusionSDF, 2023ControllableDiffusion}. However, these methods require ground truth meshes instead of point clouds during the training stage.

\section{Preliminaries: winding number and its application in surface reconstruction} \label{sec:pre}

\subsection{Inside/outside determination and surface reconstruction} \label{sec:31}

The winding number stands as a valuable tool to determine the spatial relationship between a given query point $\textbf{x} \in \mathbb{R}^3$ and a specified surface $\partial \Omega$. When $\Omega$ is an open and bounded set with a smooth boundary in $\mathbb{R}^3$, the winding number is defined as a surface integral of $\partial \Omega$ with respect to the point $\textbf{x}$:
\begin{equation}
\label{eg:gauss}
\chi(\textbf{x}) = \int_{\partial \Omega} K(\textbf{x}, \textbf{y}) \cdot \vec{N}(\textbf{y})\
\mathrm{d} S(\textbf{y}),  \ \ \ \  K(\textbf{x}, \textbf{y})=-\frac{(\textbf{x} - \textbf{y})}{4\pi ||\textbf{x}-\textbf{y}||^{3}}.
\end{equation}
Here, $\vec{N}(\textbf{y})$ represents the outward unit normal vector at point $\textbf{y}$, and $\mathrm{d} S(\textbf{y})$ denotes the surface element. $K(\textbf{x}, \textbf{y})$ is a vector-valued function with a range dimension of $3$. It is the gradient of the fundamental solution of the 3D Laplace equation.

The resulting value of the integral $\chi(\textbf{x})$ serves as the indicator function of the surface $\partial \Omega$, determining whether the query point $\textbf{x}$ lies inside, outside, or on the surface. Specifically, $\chi(\textbf{x})=0$ for $x\in \mathbb{R}^3  \setminus \overline{\Omega}$ and $\chi(\textbf{x})=1$ for $\textbf{x} \in \Omega$. When $x$ lies on the boundary, the integral is treated as a limit by integrating over $\partial \Omega  \setminus (\partial \Omega \cap B_{\epsilon}(\textbf{x}))$ and taking $\epsilon \to 0$, where $B_{\epsilon}(\textbf{x})$ is a ball centered at $\textbf{x}$ with radius $\epsilon$. As introduced in~\cite{1995PDEbook}, this limit is evaluated to $1/2$.

Once the indicator function for a closed surface in the ambient space is determined using Eq.~\eqref{eg:gauss}, the surface can be reconstructed by extracting the iso-surface with a value of $1/2$. Techniques such as GR~\cite{2019GR} and fast winding~\cite{2018FastWinding} utilize the discrete form of Eq.~\eqref{eg:gauss} for surface reconstruction from oriented point clouds. The indicator function of the input is computed directly using the following discrete winding number formulation:
\begin{equation}
\label{eq:dis_gauss}
\chi(\textbf{x}) = \sum_{i=1}^{N} {K(\textbf{x}, \textbf{y}_i) \cdot \mu_{i}}, \ \ \ \ \mu_{i} = a_{i} \textbf{n}_{i},
\end{equation}
where $\textbf{n}_{i}$ is the outward unit normal of the point $\textbf{y}_{i}$, ${a}_{i}$ is the discrete surface element computed using the geodesic Voronoi area associated with $\textbf{y}_{i}$, and $\mu_{i}$ is a 3-dimensional vector representing the multiplication of the normal and the discrete surface element. $\mu_{i}$ can be understood as the surfel mentioned in~\cite{2000Surfel}. 

Note that the integrand becomes singular as $\textbf{x}$ approximates $\textbf{y}_{i}$. This situation should be carefully considered, particularly when $\textbf{x}$ approaches the surface. Consequently, GR~\cite{2019GR} modifies the integrand of Eq.~\eqref{eq:dis_gauss} as follows:
\begin{equation}
\label{equation_modified}
\tilde{K}(\textbf{x}, \textbf{y}_i)=\left\{
\begin{aligned}
& -\frac{(\textbf{x} - \textbf{y}_i)}{4\pi ||\textbf{x}-\textbf{y}_i||^{3}} & ||\textbf{x}-\textbf{y}_i|| \geq w(\textbf{x})  \\
&-\frac{(\textbf{x} - \textbf{y}_i)}{4\pi w(\textbf{x})^{3}} & ||\textbf{x}-\textbf{y}_i|| < w(\textbf{x}).
\end{aligned} \right.
\end{equation}
In this modification, the width coefficient $w(\textbf{x})$ is determined by a specific formulation outlined in GR. The use of the new kernel function $\tilde{K}$ significantly enhances the algorithm performance for surface reconstruction. In the experimental section, we maintain a fixed value of $w(\textbf{x})$ as $0.04$.

\subsection{Surface reconstruction for unoriented point sets}
The reconstruction algorithm described in the previous section for GR requires consistently oriented normals as inputs. The reason is that the winding number formula relies on these normals to determine the spatial relationship. A recent approach PGR~\cite{2022PGR} eliminates the need for normals by treating the surfels $\mu_{i}$ in Eq.~\eqref{eq:dis_gauss} as unknowns and solves a linear system based on the on-surface constraints -- the point cloud is sampled from the surface, indicating that $\chi(\textbf{x})$ should equal to $1/2$ at all the sample points $P=\{\textbf{p}_1, \textbf{p}_2, ..., \textbf{p}_N\}$. Here, $N$ denotes the number of points. This observation leads to $N$ equations:
\begin{equation}
\label{eq:pgr_sample}
\chi(\textbf{p}_j) = \frac{1}{2}, \ \ j=1,2,...,N.
\end{equation}
We can derive the following set of linear equations by combining Eq.~\eqref{eq:dis_gauss},~\eqref{equation_modified}, and~\eqref{eq:pgr_sample}:
\begin{equation}
\label{eq:combine_qe}
\sum_{i=1}^{N} {\tilde{K}(\textbf{p}_j, \textbf{p}_i) \cdot \mu_i} = \frac{1}{2}, \ \ j=1,2,...,N.
\end{equation}
Each $\mu_{i}$ is a 3-dimensional vector with $\mu_{i} = (\mu_{i1}, \mu_{i2}, \mu_{i3})^{\top}$. These equations can be expressed in matrix form as follows:
\begin{equation}
\label{eq:linear_eq}
\textbf{A}(P)\mu=\textbf{b},
\end{equation}
where $\mu=(\mu_{11},\mu_{12},\mu_{13},...,\mu_{N1},\mu_{N2},\mu_{N3})^{\top}$, $\textbf{b}=\frac{1}{2} \vec{1}$ is a vector where all elements are $1/2$, and $\textbf{A}(P)$ is a $N \times 3N$ matrix with
\begin{equation}
\label{eq:A1}
[\textbf{A}(P)]_{j, 3i+k} = [\tilde{K}(\textbf{p}_j, \textbf{p}_i)]_{k}.
\end{equation}
The matrix $\textbf{A}(P)$ is solely dependent on the sample points $P=\{\textbf{p}_1,\textbf{p}_2,..,\textbf{p}_N\}$. In PGR, this set of linear equations is transformed into a linear least squares problem. It is worth noting that this problem is underdetermined, with $N$ equations and $3N$ unknowns. Moreover, the matrix $\textbf{A}^{\top} \textbf{A}$ typically has a large condition number. PGR introduces a regularization term $\textbf{R}(P)$ to alleviate this issue. The resulting optimization problem is:
\begin{equation}
\label{eq:pgr}
\min_{\mu} ~ \frac{1}{N} (||\textbf{A}(P)\mu - \textbf{b}||^{2} + \alpha \mu^{\top} \textbf{R}(P) \mu),
\end{equation}
with
\begin{equation}
\label{eq:rp1}
\textbf{R}(P) = \mathrm{diag}(\textbf{A}(P)^{\top} \textbf{A}(P)).
\end{equation}
Here, ``diag'' represents a matrix that retains only its diagonal elements. The parameter $\alpha$ is user-specific, and we set it to $0.5$ in the experimental section. By solving this optimization problem, we can obtain the surfels $\mu_{i}$ with globally consistent orientations, thereby allowing us to reconstruct the surface using the GR method in Section~\ref{sec:31}.
\section{Winding clearness}\label{sec:sec4}
\begin{figure}[htb]
  \centering
  \includegraphics[width=1.0\linewidth]{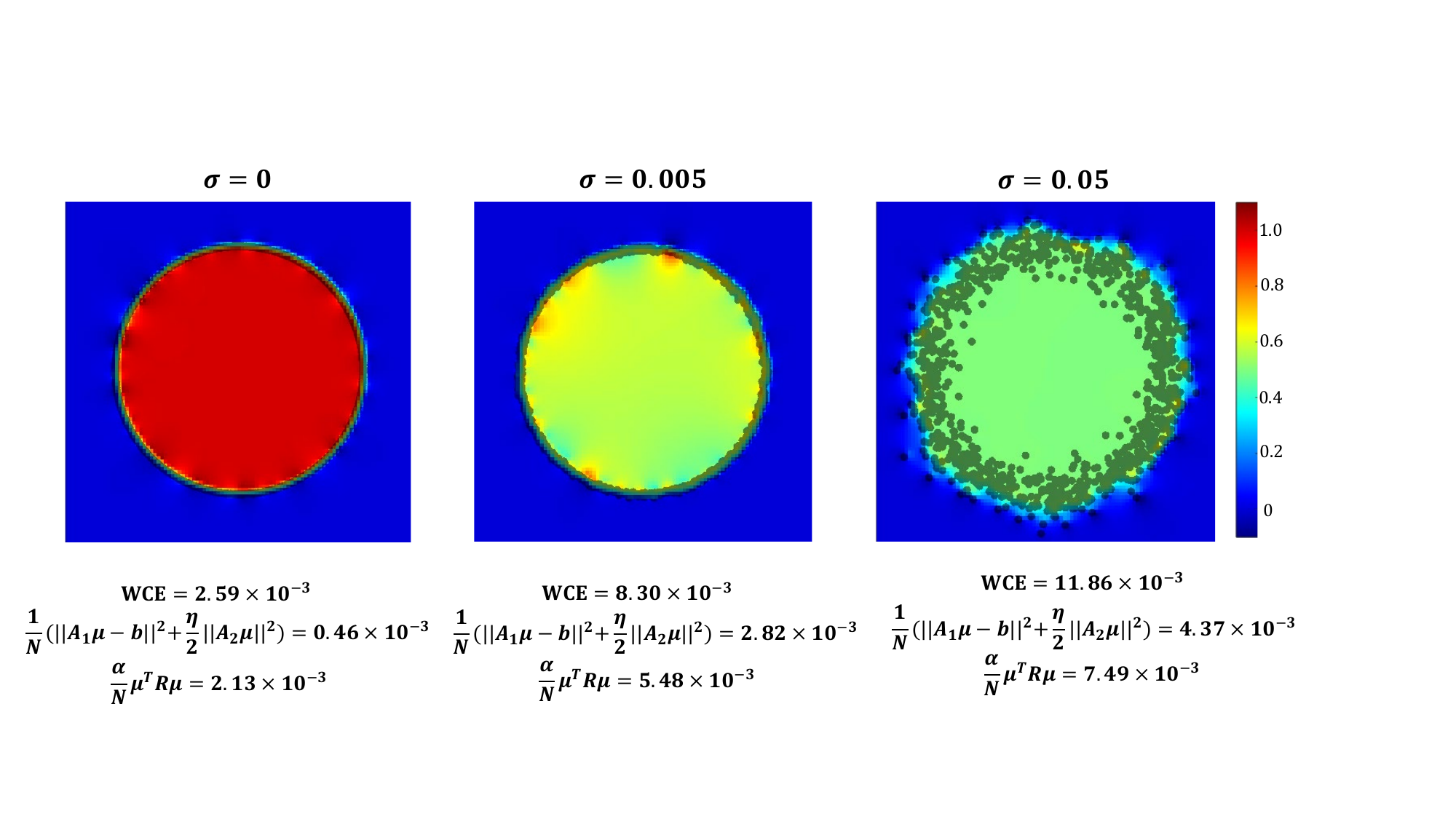}
  \caption{\label{fig:Figure1}
           Winding number field formed by 2D circles within noise of different standard deviations. We annotate the winding clearness error (WCE) and the values of its constituents. The black dots represent the sample points.}
\end{figure}

In this section, we aim to expand the applicability of the winding number theory to a broader range of point geometry problems. The algorithm discussed in the preceding section is mainly focused on optimizing the surfels to ensure that the implicit field closely approximates $1/2$ at the sampling points. Intuitively, a high-quality point cloud, such as one with minimal noise, typically displays clear internal and external relationships in the resulting winding number field, as depicted in Figure~\ref{fig:Figure1}. Consequently, the optimal value of Eq.~\eqref{eq:pgr} will be small in this case. Hence, we introduce the concept of \emph{winding clearness}, which represents the difference between the actual values and the expected values of the winding number field at specific points. Our observations also indicate that the winding number should equal to zero outside the surface. In addition to using the value of Eq.~\eqref{eq:pgr} as the measure of the winding clearness, we establish a bounding box larger than the original object and sample $2N$ points $Q=\{\textbf{q}_1,\textbf{q}_2,...,\textbf{q}_{2N}\}$ on this box. These boundary points are predetermined and remain unchanged throughout the optimization. The expected indicator function values for these $2N$ points should be zero:
\begin{equation}
\label{eq:pgr_sample2}
\chi(\textbf{q}_j) = 0, j=1,...,2N.
\end{equation}
These supplementary constraints lead to:
\begin{equation}
\label{eq:linear_eq2}
\textbf{A}_2(P)\mu=\vec{0}, \ \ \ \ [\textbf{A}_2(P)]_{j, 3i+k} = [\tilde{K}(\textbf{q}_j, \textbf{p}_i)]_{k}.
\end{equation}
The boundary points $Q$ will not be optimized during the algorithm in the next section. Accordingly, we do not express $Q$ as the variables. By combining Eq.~\eqref{eq:pgr_sample} and~\eqref{eq:pgr_sample2}, and denoting $\textbf{A}(P)$ in Eq.~\eqref{eq:linear_eq} as $\textbf{A}_1(P)$, we can express the modified objective function as follows:
\begin{equation}
\label{eq:min}
\min_{\mu} ~ f(P, \mu),
\end{equation}
where
\begin{equation}
\label{eq:fmu}
f(P, \mu) = \frac{1}{N} (||\textbf{A}_{1}(P)\mu - \textbf{b}||^{2} + \frac{\eta}{2}  ||\textbf{A}_{2}(P)\mu||^{2} + \alpha \mu^{\top} \textbf{R}(P) \mu).
\end{equation}
In the above equation, $\mu$ also represents the combined vector of surface element multiplies the normal, which is identical to Eq.~\eqref{eq:linear_eq}. $\eta$ is a user specific parameter to control the contribution of the bounding box constraint. The expression for the regularization term $\textbf{R}(P)$ is modified as follows:
\begin{equation}
\label{eq:rp2}
\textbf{R}(P) = \mathrm{diag}(\textbf{A}_{1}(P)^{\top} \textbf{A}_{1}(P) + \frac{\eta}{2} \textbf{A}_{2}(P)^{\top} \textbf{A}_{2}(P)). 
\end{equation}
The minimization problem of Eq.~\eqref{eq:min} can be solved by $\nabla_{\mu} f(P, \mu)=0$, which leads to 
\begin{equation}
\label{eq:mu}
\small
\mu(P) = ({\textbf{A}_{1}(P)}^{\top} \textbf{A}_{1}(P) + \frac{\eta}{2}{\textbf{A}_{2}(P)}^{\top} \textbf{A}_{2}(P) + \alpha \textbf{R}(P))^{-1} ({\textbf{A}_{1}(P)}^{\top}\textbf{b}).
\end{equation}
We will demonstrate the influence of the bounding box constraint through ablation studies, evaluating both the parameter $\eta$ and the bounding box location in Appendix D.4. In Eq~\eqref{eq:mu}, $\mu(P)$ only depends on the point positions $P$. Moreover, as $\textbf{A}_1, \textbf{A}_2, \textbf{R}$ are all smooth functions of $P$ and matrix inversion is differentiable, $\mu(P)$ is a differentiable function of $P$. The winding clearness error $W(P)$ is then defined as the minimum value of $f$:
\begin{equation}
\label{eq:wp}
W(P)=f(P, \mu(P))=\frac{1}{N} (\textbf{b}^{\top}\textbf{b}-\textbf{b}^{\top}\textbf{A}_{1}(P)\mu(P)),
\end{equation}
which is also a differentiable function of $P$. This important property enables us to employ winding clearness in point cloud processing pipelines that require differentiability, including both optimization-based and learning-based manner. This will be discussed in Section~\ref{sec:Optimization}.

A lower value of the function $W(P)$ signifies a better winding clearness. In Table~\ref{Table0}, we present the winding clearness error for point clouds of a 2D circle composed of $1000$ points with a radius of $0.5$. We add randomized Gaussian noise with varying standard deviations $\sigma$ ranging from $0$ to $0.05$ to the point cloud. The bounding box $Q$ is defined as a cubic region with the coordinate range as $[-0.7, 0.7]$ and $\eta$ is set to $50.0$ for robustness against various levels of noise. In the 2D scenario, the winding number theory remains relevant. However, the integrand should be replaced with the directional derivative of the fundamental solution of the 2D Laplace equation rather than the 3D counterpart. The results indicate that the winding clearness error increases as the noise level rises. Figure~\ref{fig:Figure1} showcases the winding number field generated by varying noise standard deviations, along with the annotation of the winding clearness error (WCE) and its constituent values. In the case of point clouds with substantial noise, achieving surfels that bring the implicit field of all the sampling points to $1/2$ becomes increasingly challenging. Although the surfels solved in high noise conditions can result in a considerable number of points approaching $1/2$, as illustrated in $\sigma=0.05$ of Figure~\ref{fig:Figure1}, the surfels become large in this case, resulting in a substantial increase in the regularization term $\mu^{\top} \textbf{R}(P) \mu$. Further explanation of the regularization term can be found in Appendix A of the supplementary material.

\begin{table}[t]
\centering
\caption{The winding clearness error (WCE) of a circle composed of $1000$ points with a radius of $0.5$. We add randomized Gaussian noise with varying standard deviations $\sigma$ ranging from $0$ to $0.05$. The WCE value is multiplied by $10^{3}$.}
\label{Table0}
\begin{tabular}{cccccccc}
\hline
$\sigma$ & $0$ & $0.001$ & $0.002$ & $0.005$ & $0.01$ & $0.02$ & $0.05$\\
\hline
WCE & $2.59$ & $3.29$ & $4.41$ & $8.30$ & $8.94$ & $9.62$  & $11.86$\\

\hline
\end{tabular}
\label{table_MAP}
\end{table}

\section{Optimization of winding clearness}
\label{sec:Optimization}

The winding clearness error $W(P)$ can be calculated for a raw point cloud $P$ using Eq.~\eqref{eq:mu} and~\eqref{eq:wp}. It is reasonable to assume that improving the winding clearness, indicated by reducing the value of $W(P)$, will have a positive influence on the overall quality of the point cloud. The computation of $W(P)$ is fully differentiable because it mainly involves linear algebraic operations. Consequently, we propose to utilize $W(P)$ as the loss function for point cloud optimization. The value of $W(P)$ can be directly back-propagated to the coordinates of all points, which holds the potential to enhance the quality of the point cloud by reducing the noise. In this regard, we present two approaches: 1) Directly optimizing the winding clearness of an individual point cloud (Section~\ref{sec:51}), and 2) Incorporating winding clearness as a geometric constraint in diffusion-based generation (Section~\ref{sec:52}).

\subsection{Direct optimization of single point cloud}\label{sec:51}
We can directly utilize the gradient decent approach to optimize the winding clearness of a given point cloud without a deep neural network. To ensure that the resulting point cloud does not deviate excessively from the original one (denoted as $P_{0}$), we incorporate an $L_{2}$ regularization term to penalize the distance between the result and the original point cloud. The loss function is defined as follows:
\begin{equation}
\label{eq:optimize}
\mathrm{Loss}(P) = W(P) + \frac{\lambda}{N} ||P - P_{0}||^{2},
\end{equation}
where $\lambda$ is a user specific parameter and $N$ represents the number of points. We will conduct experiments for different $\lambda$ values in Appendix D.5. The calculation of $W(P)$ involves obtaining $\mu(P)$ by computing an inverse matrix as shown in Eq.~\eqref{eq:mu}, which can be address by solving a linear system. Specifically,
\begin{align}
\label{eq:torch_solve}
&\tilde{\textbf{A}}={\textbf{A}_{1}(P)}^{\top} \textbf{A}_{1}(P) + \frac{\eta}{2}{\textbf{A}_{2}(P)}^{\top} \textbf{A}_{2}(P) + \alpha \textbf{R}(P), \\
&\tilde{\textbf{b}}={\textbf{A}_{1}(P)}^{\top}\textbf{b}.
\end{align}
Then, $\mu$ can be obtained by solving
\begin{equation}
\label{eq:torch_solve_mu}
\tilde{\textbf{A}} \mu = \tilde{\textbf{b}}.
\end{equation}
The solving process ensures differentiability whether utilizing LU decomposition~\cite{2003Strang} or the conjugate gradient method~\cite{1952CG}, as both involve basic linear algebraic calculations. We note that PyTorch~\cite{2019pytorch} provides a numerically stable and differentiable function, \emph{torch.linalg.solve}, which solves the linear system with back-propagation. Therefore, we directly employ this function along with the gradient-based Adam optimizer~\cite{2014Adam}. After solving for $\mu$ using Eq.~\eqref{eq:torch_solve_mu}, we can calculate $W(P)$ by inserting $\mu(P)$ to $f(P, \mu)$ in Eq.~\eqref{eq:fmu}. The derivative ${\mathrm{d} W(P)}$/${\mathrm{d} P}$ can then be computed utilizing the chain rule and back-propagation mechanism. The loss function directly facilitates the adjustment of the point coordinates, leading to improved winding clearness.

\begin{figure}[htb]
  \centering
  \includegraphics[width=1.0\linewidth]{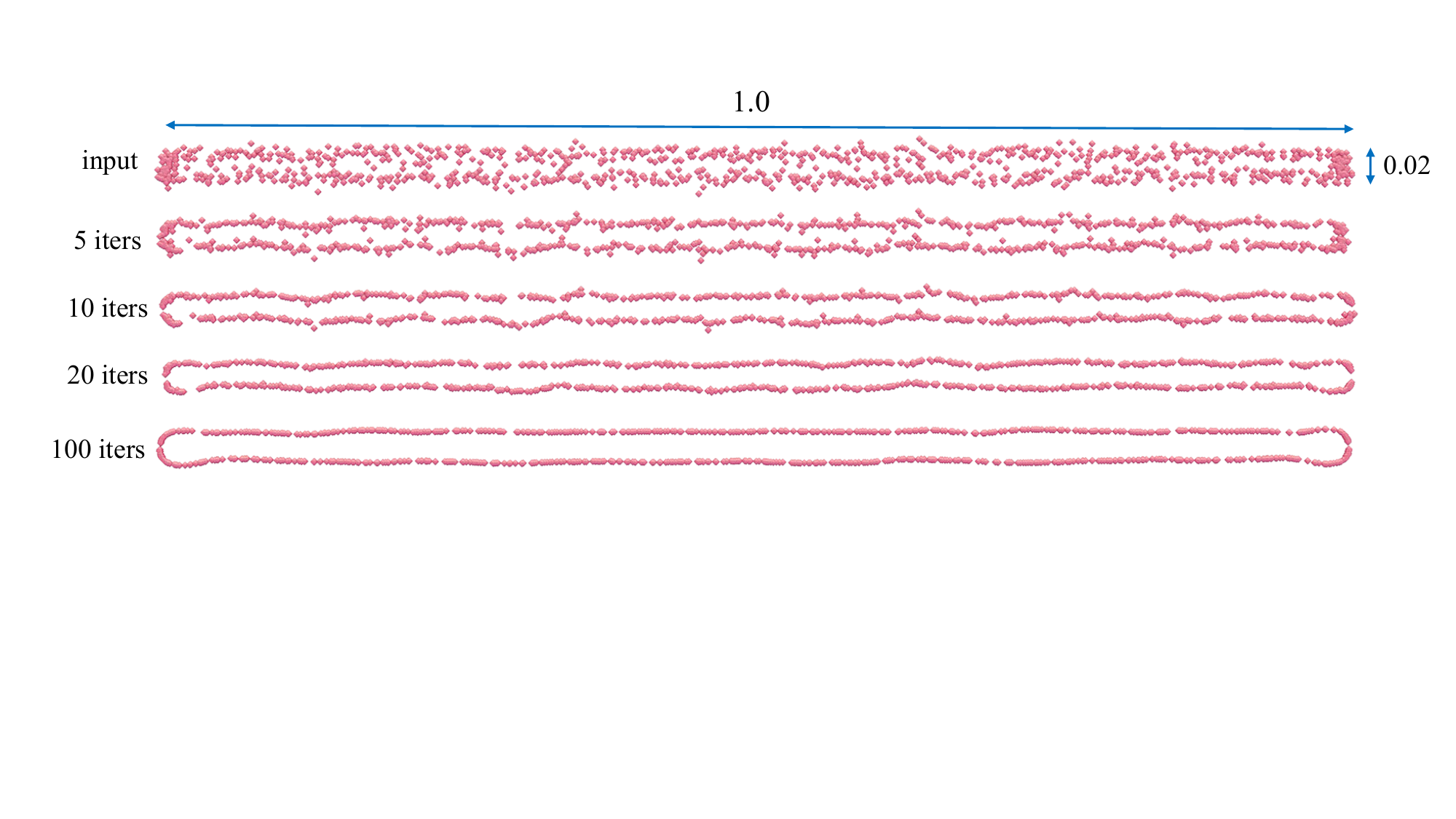}
  \caption{\label{fig:Figure7}
           Iterative process of our method applied to a noisy and thin point set.}
\end{figure}

Figure~\ref{fig:Figure7} depicts a 2D example featuring a thin structure to demonstrate the effectiveness of our method. The input consists of $1000$ points sampled from an slender rectangle with a major axis of $1.0$ and a minor axis of $0.02$. Subsequently, per-point Gaussian noise of the standard deviation $0.004$ is added to the point set. We optimize the point cloud through Eq.~\eqref{eq:optimize}. After $100$ iterations, the resulting point cloud shows reduced noise, and the rectangle structure is well preserved. This experiment underscores the ability of our method to handle this situation. Although some recent learning-based methods also recognize this challenge~\cite{2024SharpNeural}, they still depend on training data to learn the shape priors. In contrast, our method optimizes the winding clearness through a non-data-driven approach.

For stable numerical solution of the equation system in Eq.~\eqref{eq:torch_solve_mu}, we require the matrix $\tilde{\textbf{A}}$ to be non-singular and to have a moderate condition number. When these conditions are not met, applying a pseudo-inverse operation provides greater numerical stability. However, the set of non-invertible matrices has a measure of zero within the set of all matrices of the same dimension. Moreover, the regularization term $\textbf{R}(P)$ further reduces both the probability of encountering singular matrices and the occurrence of cases with excessively large condition numbers. In all our experiments conducted so far, the direct solution using the LU decomposition-based method (e.g. torch.linalg.solve) has proven to be numerically robust.

\subsection{Winding clearness as geometric constraint in diffusion-based generation}\label{sec:52}

In addition to optimizing the point cloud directly using Eq.~(\ref{eq:optimize}), the winding clearness error $W(P)$ can also be considered as a geometric constraint and applied in various learning-based point processing tasks. In this section, we propose to incorporate this consideration with the diffusion-based 3D generative model. The Denoising Diffusion Probabilistic Model (DDPM)~\cite{2020DDPM} has gained significant popularity and is utilized in various 3D point cloud generation tasks~\cite{2021DPM, 2021PVD, 2022LION, 2023FastDiffusion}. The current loss functions of these methods primarily focus on ensuring the similarity between the generated data and the original dataset, but lack a specific term for constraining the geometric properties. Our method provides a solution for this gap by optimizing the winding clearness of the generated results through a joint training strategy.

We choose two generative frameworks as baselines: one based on standard DDPM and the other based on velocity flow. In the following sections, we will introduce how to incorporate winding clearness as a geometric constraint in PVD~\cite{2021PVD}, which adopts a standard DDPM process and demonstrates competitive performance in point cloud generation. The methodology in the velocity flow will be discussed in Appendix C.

The diffusion process of PVD involves two steps: a noise adding process $q(\textbf{x}_{0:T})$ and a denoising process $p_{\theta}(\textbf{x}_{0:T})$ performed over $T$ steps:
\begin{align}
& q(\textbf{x}_{0:T}) = q(\textbf{x}_{0})\prod_{t=1}^{T} q(\textbf{x}_{t}|\textbf{x}_{t-1}),\\
& p_{\theta}(\textbf{x}_{0:T})=p(\textbf{x}_{T})\prod_{t=1}^{T}p_{\theta}(\textbf{x}_{t-1}|\textbf{x}_{t}),
\end{align}
where $q(\textbf{x}_{0})$ signifies the data distribution, and $p(\textbf{x}_{T})$ denotes an initial Gaussian prior. The sequence $\textbf{x}_T, \textbf{x}_{T-1}, ..., \textbf{x}_0$ illustrates the denoising process within the diffusion model. $\theta$ corresponds to the network parameters. The Point-Voxel CNN~\cite{2019PVCNN} is chosen as the backbone network of PVD, specifically corresponding to $\epsilon_{\theta}$ in Eq.~\eqref{eq:epsilon} below.

During the training process, PVD learns the marginal likelihood and maximizes the variational lower bound similar to traditional DDPM, which can be expressed as follows:
\begin{equation}
\label{eq:log}
\mathop{\max}_{\theta} E_{q(\textbf{x}_0)}[\mathrm{log} \ p_{\theta}(\textbf{x}_0)].
\end{equation}
After reparameterization and rigorous mathematical deduction  (as described in \cite{2020DDPM} and \cite{2021PVD}), the training objective function is transformed to:
\begin{equation}
\label{eq:epsilon}
\min ||\epsilon - \epsilon_{\theta}(\textbf{x}_t, t)||^{2}, \epsilon \sim \mathcal{N}(0,\mathcal{I}).
\end{equation}
The training process aims to make the values $\epsilon_{\theta}$ approach a normal distribution for specific variables.

The test process of PVD involves generating a point cloud $\tilde{\textbf{x}}_0$ from a randomly sampled initial Gaussian prior $\tilde{\textbf{x}}_T$ and a series of Gaussian noise $\textbf{z}_{t}$. The generation process can be described as follows:
\begin{equation}
\label{eq:test_seq}
\tilde{\textbf{x}}_{t-1} = \frac{1}{\sqrt{\alpha_{t}}}(\tilde{\textbf{x}}_t - \frac{1-\alpha_{t}}{\sqrt{1-\tilde{\alpha}_{t}}} \epsilon_{\theta}(\tilde{\textbf{x}}_{t}, t)) + \sqrt{\beta_t} \textbf{z}_{t}.
\end{equation}
Here, all $\alpha_t$, $\tilde{\alpha}_t$ and $\beta_{t}$ values are specific or computed parameters in the DDPM, $\textbf{z}_{t} \sim \mathcal{N}(0,\mathcal{I})$, and $\epsilon_{\theta}$ is the same as Eq.~\eqref{eq:epsilon}. This generation process can be represented as:
\begin{equation}
\label{eq:test_process}
\tilde{\textbf{x}}_{0} = G(\tilde{\textbf{x}}_{t}, \theta, \textbf{z}, \alpha, \beta).
\end{equation}
where $\textbf{z}$ represents all $\textbf{z}_{t}$ values, $\alpha$ denotes all $\alpha_{t}$ and $\tilde{\alpha}_{t}$ values, and $\beta$ indicates all $\beta_{t}$ values. We do not provide detailed explanations for all these parameters in our work due to space constraints, more contents about DDPM can be found in ~\cite{2020DDPM} and~\cite{2021PVD}.

To incorporate winding clearness as the geometric constraint in the diffusion model, our approach utilizes a joint training strategy, which consists of a training branch and a generation branch. In the training branch, the loss function is the same as Eq.~\eqref{eq:epsilon}. In the generation branch, a reverse process described by Eq.~\eqref{eq:test_process} is executed while maintaining the propagation of the gradients.

For the generation branch, we sample $\tilde{\textbf{x}}_T$ from a Gaussian distribution and produce the output $\tilde{\textbf{x}}_0$ using Eq.~\eqref{eq:test_seq} and~\eqref{eq:test_process}. Wavy lines are used for distinction between the training and the generation branches. Subsequently, the winding clearness error of the generated point cloud $\tilde{\textbf{x}}_0$ is added to the loss function as $W(\tilde{\textbf{x}}_0)$. The total loss is then calculated as the combined sum of the losses from the training and generation branches. The joint training process is taking the gradient decent step as follows:
\begin{equation}
\label{eq:gradient}
\nabla_{\theta}{\Big(||\epsilon - \epsilon_{\theta}(\textbf{x}_t, t)||^{2} + \rho W(\tilde{\textbf{x}}_0)\Big)}.
\end{equation}
The parameter $\rho$ is user-specific and determines the balance between the training and generation branches. The batch size for the winding clearness term is $1$ for memory efficiency, while the batch size of the $\epsilon$ term  remains consistent with that of PVD, which is $16$.

Given that $W(\tilde{\textbf{x}}_0)$ serves as a fine-tuning of the original model, the gradient of the generation branch is only propagated when $t < t_{s}$. Specifically, when $t \geq t_{s}$ (where $t$ ranges from $T$ to $0$ in the denoising process), the gradients of the intermediate results $\tilde{\textbf{x}}_t$ are detached in the computation graph. Therefore, no gradients will be propagated beyond $t_{s}$. We set $T$ to $1000$ and $t_{s}$ to $10$ in our method.

The joint training strategy of Eq.~\eqref{eq:gradient} can commence from a pre-trained model and act as a fine-tuning process for the original model. In the experimental section, we demonstrate that a certain period of fine-tuning can substantially improve the quality of the generated point clouds.

\section{Experiments}\label{sec:6}
The experimental section is structured as follows. In Section~\ref{sec:6_1}, we will showcase the influence of different noise scales on the winding clearness error of point clouds with various shapes. Subsequently, we will show how the quality of point clouds can be improved by optimizing the winding clearness. The results of the single point cloud optimization will be presented in Section~\ref{sec:6_2}, and the results of diffusion-based method will be presented in Section~\ref{sec:6_3}.

\begin{figure}[htb]
  \centering
  \includegraphics[width=1.0\linewidth]{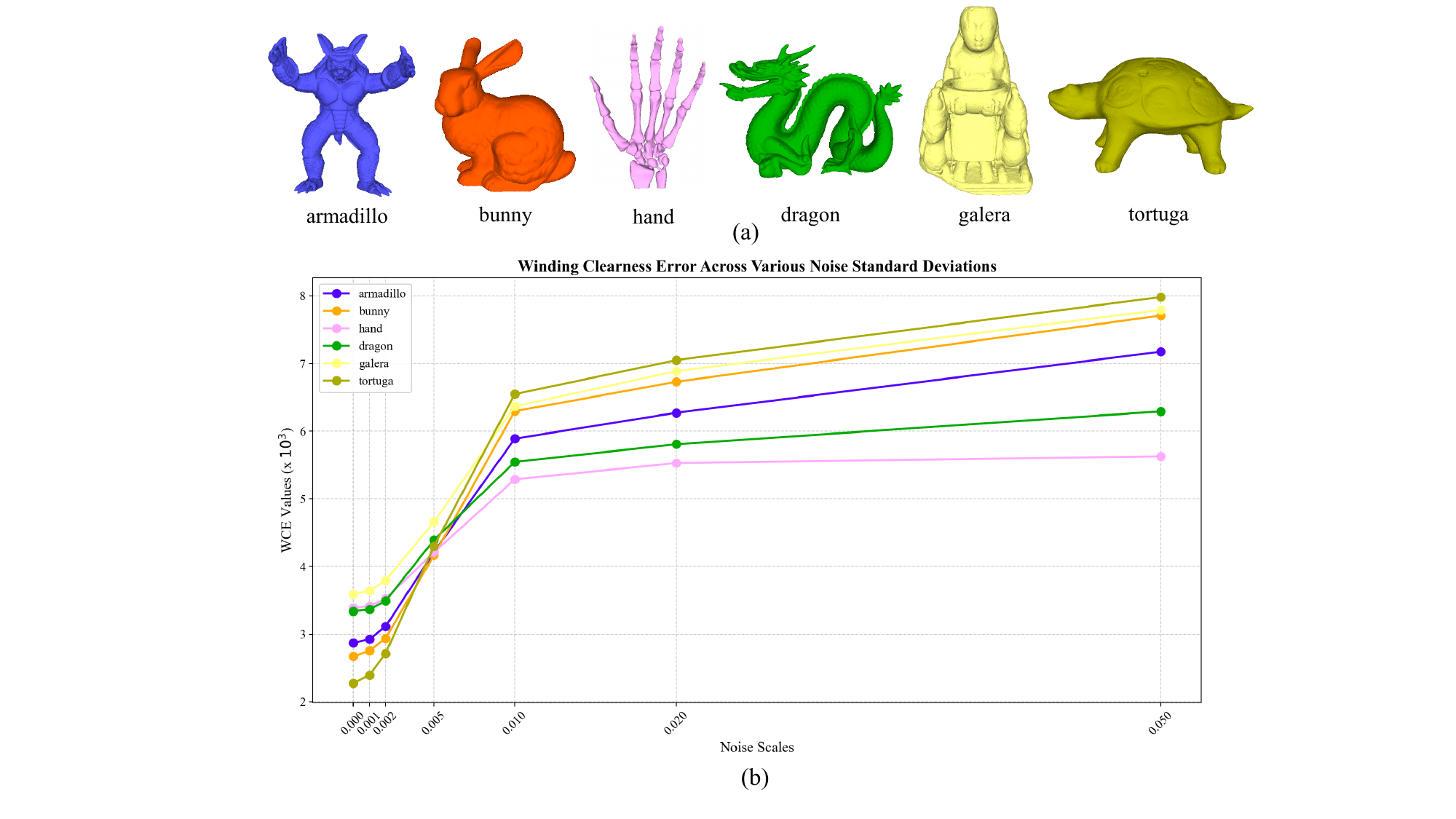}
  \caption{\label{fig:Figure9}
           Winding clearness error (WCE) of point clouds across various noise standard deviations $\sigma$ sampled from several frequently utilized shapes. The WCE values are multiplied by $10^{3}$.}
\end{figure}

\begin{figure}[htb]
  \centering
  \includegraphics[width=1.0\linewidth]{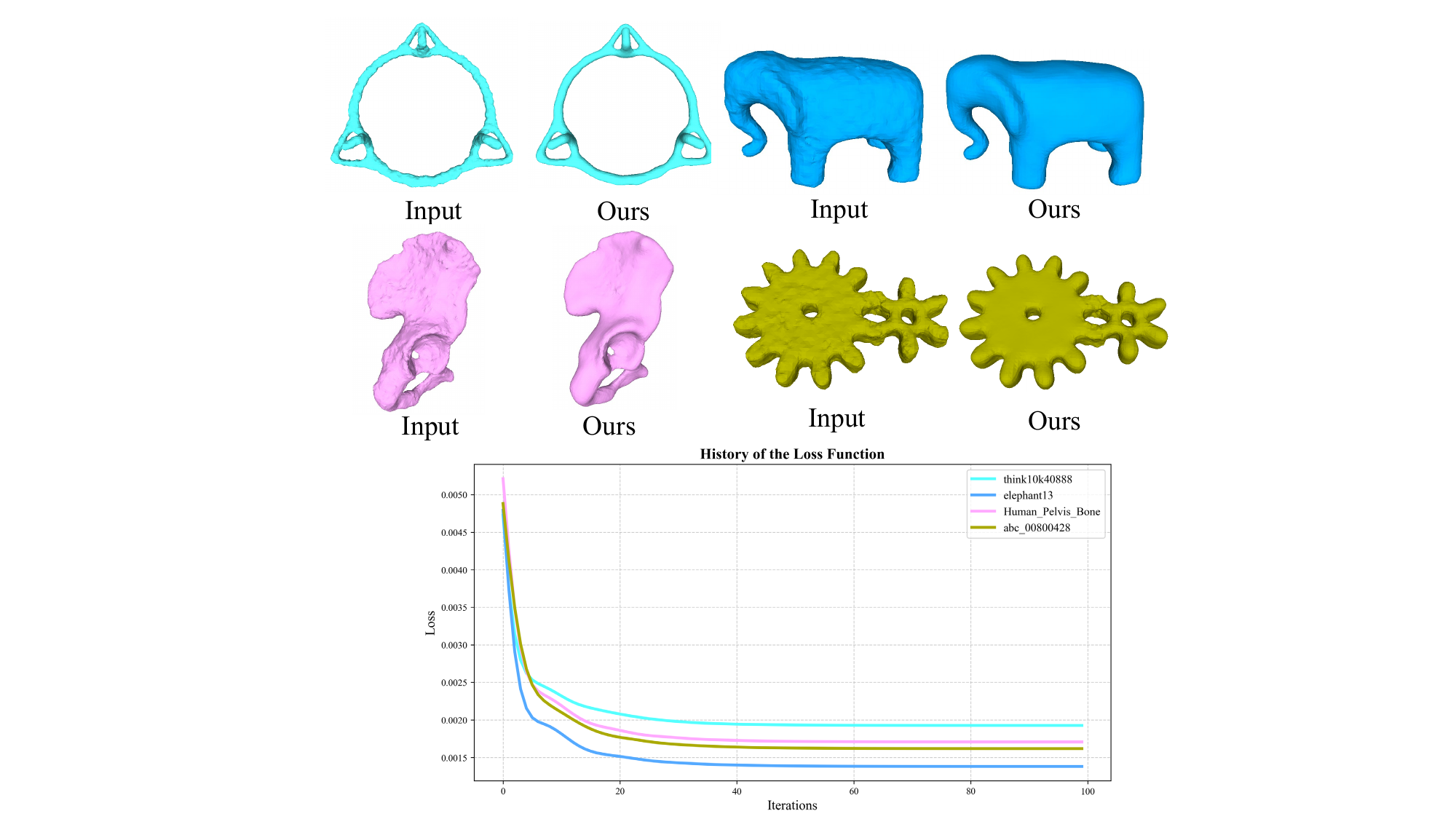}
  \caption{\label{fig:Figure5}
           Loss functions of several examples. The winding clearness can be effectively optimized, leading to a notable enhancement in the quality of the output.}
\end{figure}

\subsection{Winding clearness error of noisy inputs}\label{sec:6_1}
In Section~\ref{sec:sec4}, we have demonstrated the winding clearness error of a 2D unit circle considering varying levels of Gaussian noise. Here, our attention turns toward the evaluation of 3D shapes. In Figure~\ref{fig:Figure9}(a), we display several frequently utilized shapes in computer graphics. Each shape is normalized to a maximum axis length of $1.0$. We sample $5000$ points from each shape. Subsequently, we add randomized Gaussian noise of seven different amplitudes ($0$, $0.001$, $0.002$, $0.005$, $0.01$, $0.02$, $0.05$) for each shape. The calculation of the winding clearness error is performed utilizing Eq.~\eqref{eq:mu} and~\eqref{eq:wp}, with the bounding box $Q$ set at $-0.7$ and $0.7$, a range sufficient to encompass $3\sigma$. $\eta$ in Eq.~\eqref{eq:fmu} is set to $200.0$ for robustness against various levels of noise. Figure~\ref{fig:Figure9}(b) shows the winding clearness error of these shapes within different noise amplitudes, where we can observe a positive correlation between the winding clearness error and the noise magnitude. Therefore, we can expect that optimizing this error can help reduce the point cloud noise.

\subsection{Results of single point cloud optimization} \label{sec:6_2}

In this section, we validate the effectiveness of the single point cloud optimization approach proposed in Section~\ref{sec:51}. The loss function is defined as Eq.~\eqref{eq:optimize}, and we utilize $100$ iterations to optimize the winding clearness of each shape. Figure~\ref{fig:Figure5} shows the loss history of several shapes, along with the reconstructed mesh of the input point cloud before and after the optimization using our method. Each input comprises $5000$ sample points with Gaussian noise $\sigma=0.005$. We believe that the reconstructed mesh provides an intuitive way to reflect the noise in the point cloud. It can be observed that the loss function converges smoothly in a decreasing trend, and the quality of the output point cloud is remarkably improved.

To further validate the effectiveness of our method, we conduct qualitative comparisons with relevant approaches.  First, we will evaluate the performance of our method on several commonly used geometric shapes. Then, we will carry out experiments using a benchmark dataset compiled by a recent survey paper~\cite{2022Survey}. This dataset includes a wide-ranging collection of shapes from various repositories, such as 3DNet~\cite{3DNet2012}, ABC~\cite{2019ABC}, Thingi10K~\cite{2016Thingi10K}, and 3D Scans~\cite{20213DScans}. All the shapes in this benchmark are classified according to the reconstruction complexity. In our experiments, we specifically use the \emph{ordinary} dataset, which consists of $486$ shapes and provides a thorough evaluation of the robustness for each method. In particular, a large number of shapes in this dataset have thin structures.

\begin{figure*}[htb]
  \centering
  \includegraphics[width=0.95\linewidth]{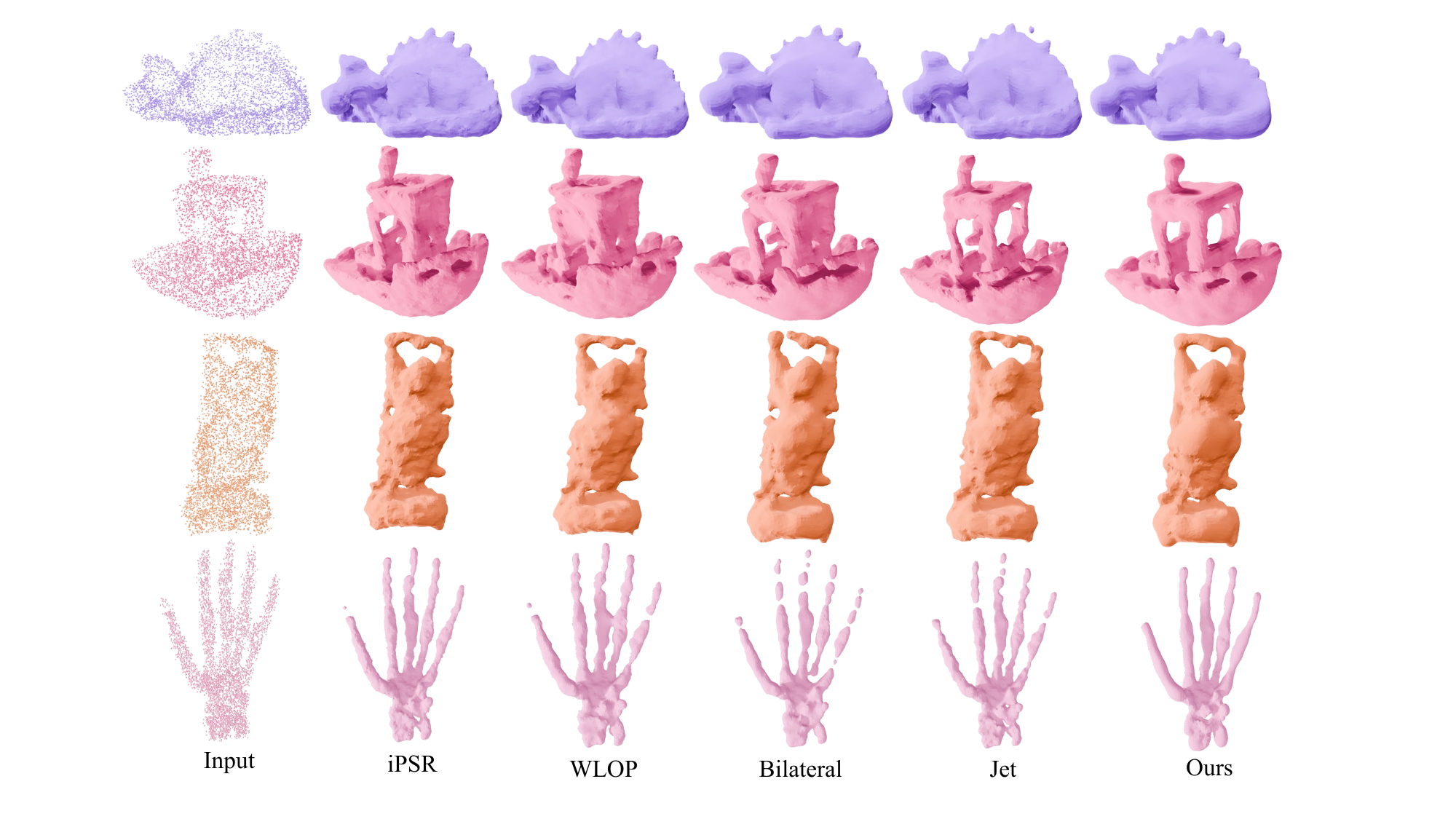}
  \caption{\label{fig:Figure4}
           Qualitative comparisons on several commonly used geometric shapes. Our method achieves well-rounded performance.}
\end{figure*}

\begin{table*}[t]
\centering
\caption{Quantitative comparisons of the reconstructed mesh quality in the ordinary dataset. We show the Chamfer distance (CD), the normal consistency (NC), and the F-score of each method. The CD values are multiplied by $10^{2}$. Our method exhibits superior performance against local feature-based methods.}
\label{table2}
\begin{tabular}{ccccccccccccc}
\toprule
\multirow{2}{*}{Model} & \multicolumn{3}{c}{WLOP} & \multicolumn{3}{c}{Bilateral} & \multicolumn{3}{c}{Jet} & \multicolumn{3}{c}{Ours} \cr
\cmidrule(lr){2-4} \cmidrule(lr){5-7} \cmidrule(lr){8-10} \cmidrule(lr){11-13}
 & CD$\downarrow$ & NC$\uparrow$ & F-score$\uparrow$ & CD$\downarrow$ & NC$\uparrow$ & F-score$\uparrow$ & CD$\downarrow$ & NC$\uparrow$ & F-score$\uparrow$  & CD$\downarrow$ & NC$\uparrow$ & F-score$\uparrow$ \\
\midrule

think10k298321 & $1.874$ & $0.777$ & $0.582$ &$1.378$ & $0.832$ & $0.674$ & $2.337$ & $0.723$ & $0.471$ & $\textbf{1.246}$ & $\textbf{0.877}$ & $\textbf{0.690}$ \\

think10k314586 & $1.391$ & $0.608$ & $0.590$ & $0.681$ & $0.849$ & $0.943$ & $0.731$ & $0.797$ & $0.898$ &  $\textbf{0.603}$ & $\textbf{0.887}$ & $\textbf{0.944}$ \\

scissors4 & $1.490$ & $0.505$ & $0.571$ & $0.786$ & $0.766$ & $0.892$ & $0.767$ & $0.742$ & $0.873$ & $\textbf{0.623}$ & $\textbf{0.827}$ & $\textbf{0.938}$ \\

think10k73183 & $1.120$ & $0.965$ & $0.813$ & $0.514$ & $0.971$ & $0.999$ & $0.637$ & $0.965$ & $0.972$ &  $\textbf{0.385}$ & $\textbf{0.986}$ & $\textbf{1.000}$ \\

saxophone2 & $0.739$ & $0.939$ & $0.900$ & $0.418$ & $0.962$ & $\textbf{0.995}$ & $0.378$ & $0.956$ & $0.992$ &  $\textbf{0.314}$ & $\textbf{0.969}$ & $\textbf{0.995}$ \\

think10k39084 & $1.110$ & $0.897$ & $0.766$ & $0.670$ & $0.942$ & $0.961$ & $0.935$ & $0.917$ & $0.829$ &  $\textbf{0.628}$ & $\textbf{0.955}$ & $\textbf{0.975}$ \\

\bottomrule

avg. of ordinary & $1.098$ & $0.868$ & $0.815$ & $1.078$ & $0.884$ & $0.845$ & $1.066$ & $0.874$ & $0.833$ &  $\textbf{0.921}$ & $\textbf{0.897}$ & $\textbf{0.886}$ \\

\hline
\end{tabular}

\label{table_MAP}
\end{table*}

\begin{figure*}[htb]
  \centering
  \includegraphics[width=0.95\linewidth]{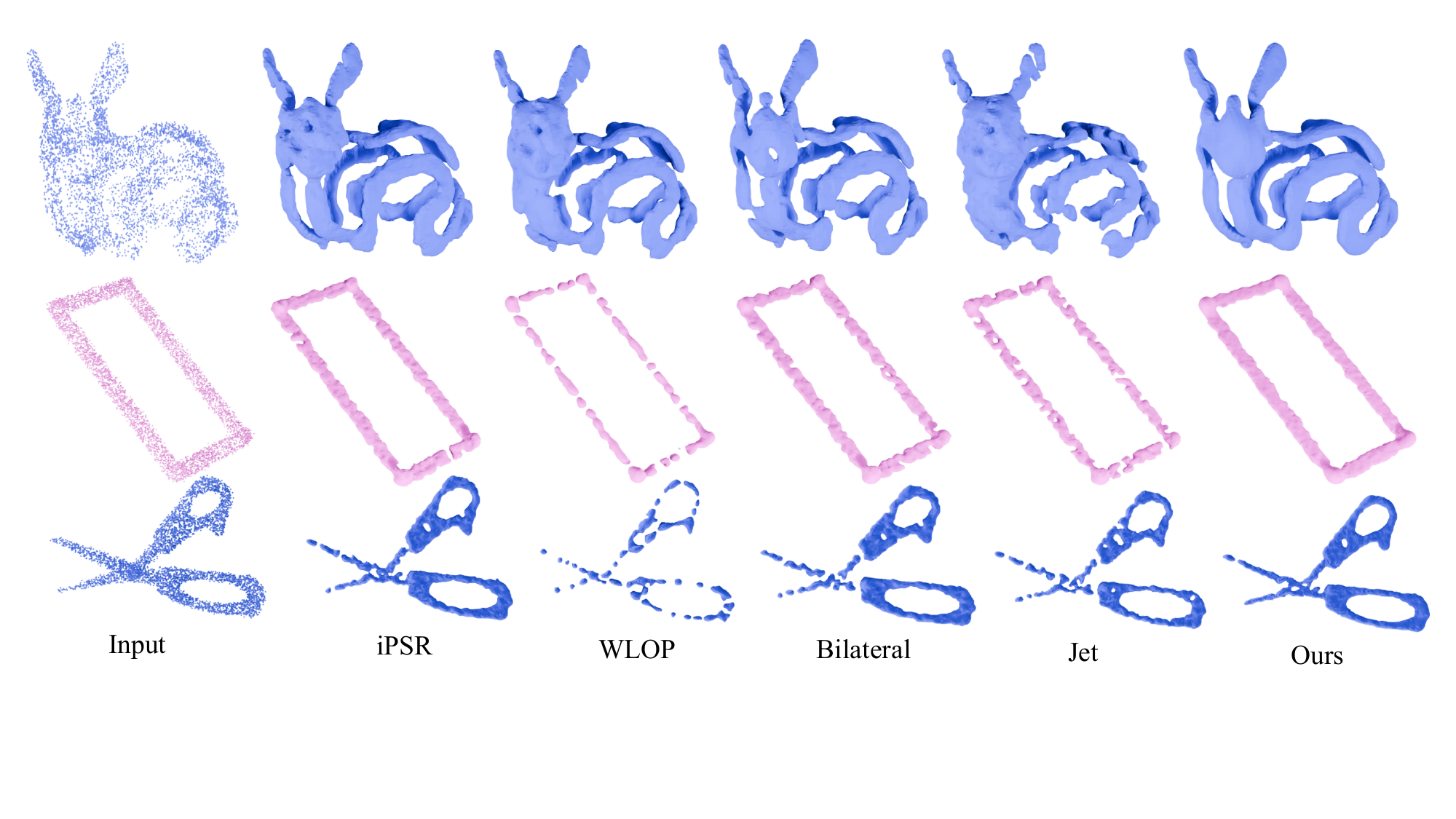}
  \caption{\label{fig:Figure4b}
           Qualitative comparisons on the ordinary dataset of sheet-structured shapes. Our method exhibits superior performance against local feature-based methods.}
\end{figure*}
\begin{figure*}[htb]
  \centering
  \includegraphics[width=0.95\linewidth]{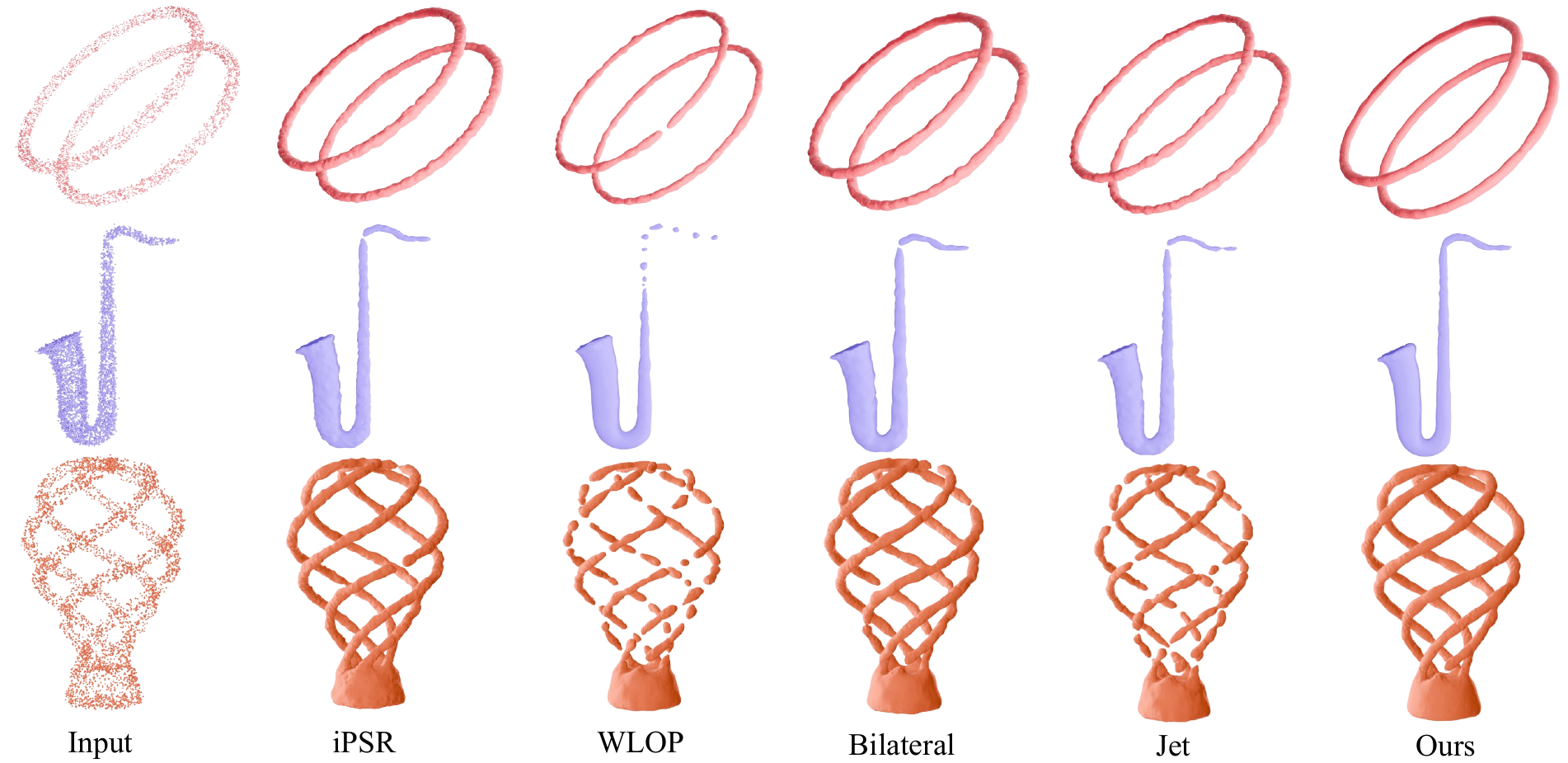}
  \caption{\label{fig:Figure4c}
           Qualitative comparisons on thin cylindrical shapes. Our method effectively manages this situation.}
\end{figure*}
\begin{figure}[htb]
  \centering
  \includegraphics[width=1.0\linewidth]{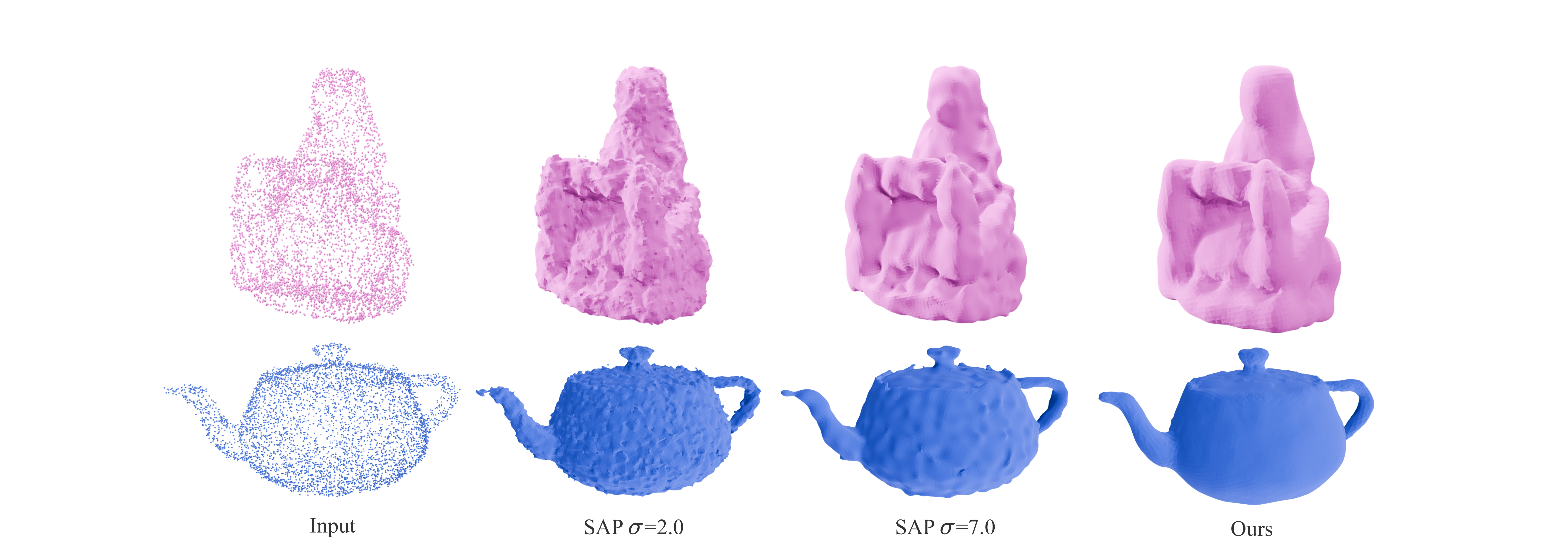}
  \caption{\label{fig:Figure8}
           Qualitative comparisons of our method with the optimization-based SAP method~\cite{2021DPSR}.}
\end{figure}

\begin{figure}[htb]
  \centering
  \includegraphics[width=0.85\linewidth]{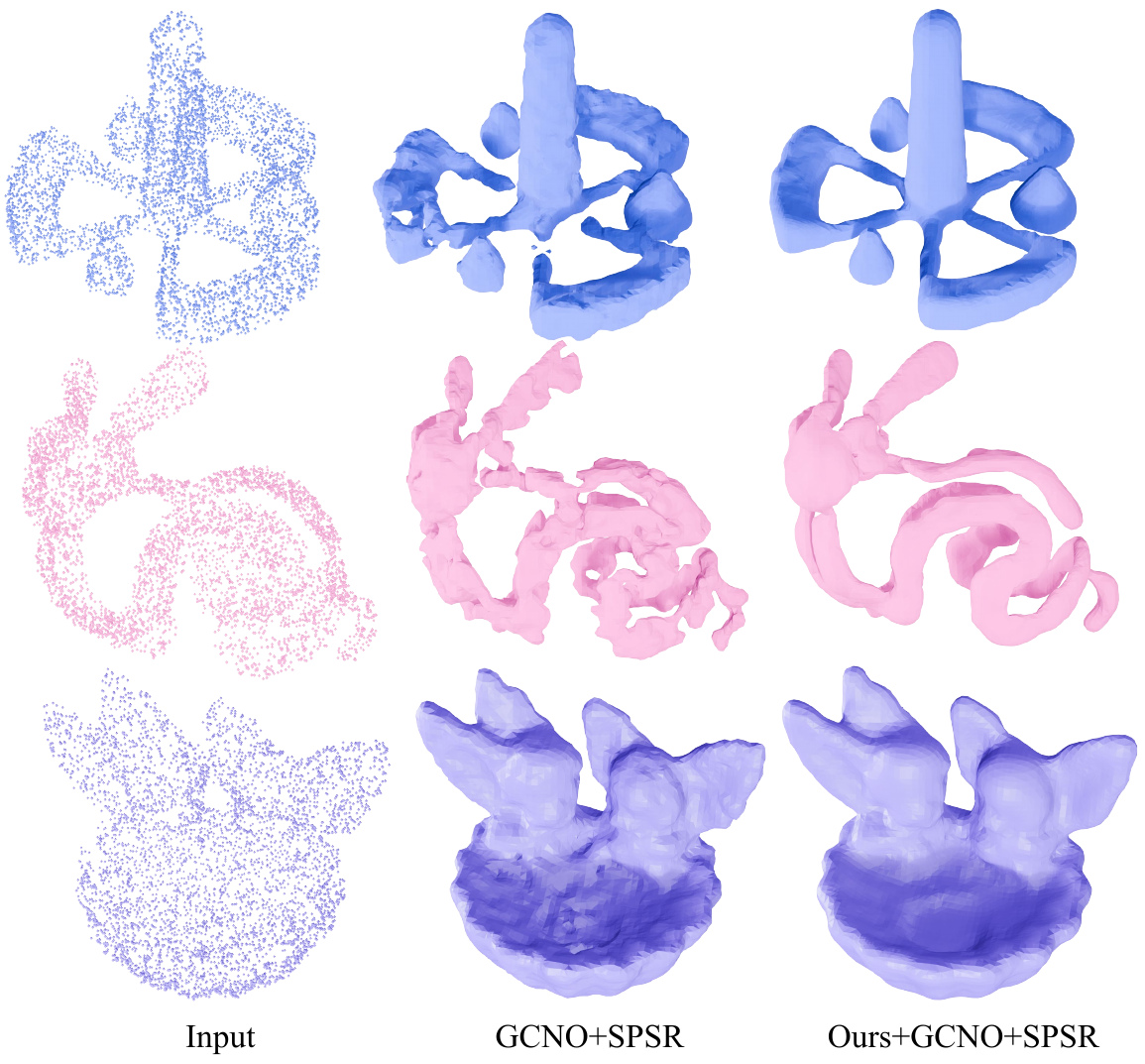}
  \caption{\label{fig:Figure10}
           Qualitative comparisons of our method with GCNO~\cite{2023GCNO}. Our method performs well on thin structures and multiple connected components.}
\end{figure}

\begin{figure*}[htb]
  \centering
  \includegraphics[width=0.99\linewidth]{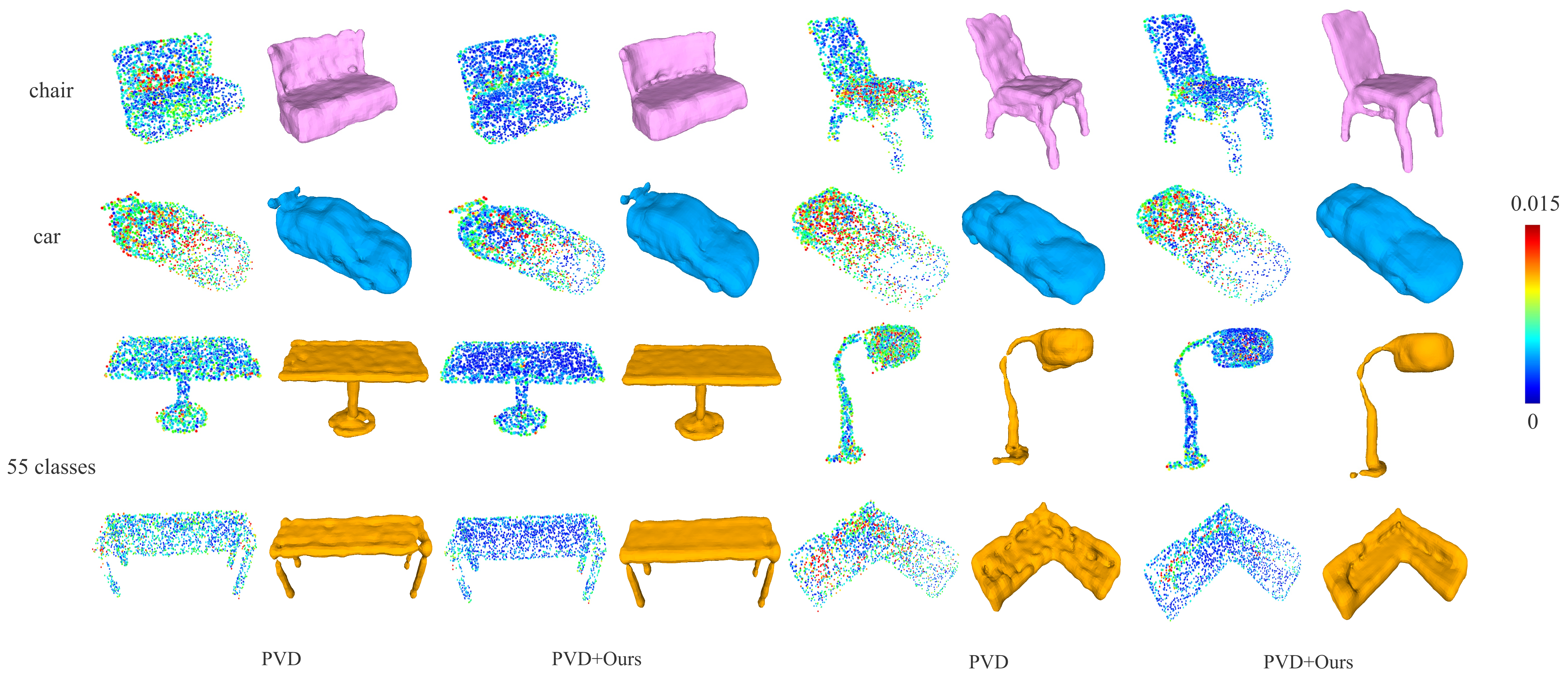}
  \caption{\label{fig:Figure6}
           Qualitative comparisons of PVD generation with and without fine-tuning by the winding clearness constraint. Our method significantly improves the quality of the generated point sets.}
\end{figure*}

We compare our method with several non-data-driven point cloud filter approaches, including bilateral smoothing~\cite{2017Bilateral}, jet smoothing~\cite{2005Jets} and WLOP~\cite{2009WLOP}. Bilateral and jet smoothing are mainly based on nearest neighbor filtering. Meanwhile, WLOP achieves local optimal projection through an iterative process. We apply the CGAL~\cite{2009CGAL} implementation of all these approaches. The bilateral implementation of CGAL also inherits the philosophy of EAR~\cite{2013EdgeAware} as illustrated in the official documentation of CGAL. In WLOP, the parameter ``representative particle number'' is set to $95\%$ of the original point set, and the parameter ``neighbor radius'' is set to $0.04$. In jet smoothing, the only parameter ``neighbor size'' is set as $15$. In bilateral smoothing, the parameters ``neighbor size'', ``sharpness angle'', and ``iters'' are set to $10$, $25$, and $5$, respectively. We use the reconstructed surface as a metric to evaluate the quality of the point cloud. For this step, an unoriented reconstruction method is required. Here, we employ iPSR~\cite{2022iPSR} with a point weight of $4.0$. We select iPSR because it can better capture the noise in thin structures. If the noise is not effectively removed, iPSR may struggle to converge and produce poor results in thin structures. Thus, iPSR offers a more intuitive demonstration of the denoising effect. Appendix B will provide detailed explanations of the relationship and conduct a comparison among our method, PGR~\cite{2022PGR}, and WNNC~\cite{2024WNNC}. Moreover, in Appendix D.1, we present a direct comparison of the denoised point sets without resorting to reconstruction.

In Figure~\ref{fig:Figure4}, we present a qualitative comparison of several commonly utilized geometric shapes among different approaches. The input datasets consist of $5000$ points, with the first two rows exhibiting randomized Gaussian noise $\sigma=0.005$ and the last two rows with $\sigma=0.01$. The parameters $\eta$ in Eq.~\eqref{eq:fmu} is set to $10.0$ and $\lambda$ in Eq.~\eqref{eq:optimize} is set to $20.0$. Meanwhile, the bounding box is established at $-0.6$ and $0.6$. The column ``iPSR'' denotes the results reconstructed from the original noisy point clouds, and the subsequent columns showcase the reconstructed meshes of point clouds generated by the corresponding denoising approaches. The results show that our method exhibits well-rounded performance. For instance, in the dragon example, it achieves better reconstruction on the jagged back. Moreover, in the remaining examples, it generates more complete and reasonable surfaces.

For the \emph{ordinary} dataset, we also sample $5000$ points for each shape and add randomized Gaussian noise $\sigma=0.005$ to the point clouds. This dataset comprises 486 shapes of various types, especially featuring numerous thin structures. The qualitative comparisons are presented in Figure~\ref{fig:Figure4b} (focusing on sheet structures) and Figure~\ref{fig:Figure4c} (focusing on thin cylindrical structures). The examples are closed manifolds with thin structures. Traditional methods may smooth out the original thickness and fail to produce satisfactory reconstructed meshes. In contrast, our method adopts a global perspective using the winding number, leading to better performance.

We also provide quantitative comparisons in terms of the two-way Chamfer distance (CD), the symmetric normal consistency (NC) and the F-score in Table~\ref{table2} with the ground truth mesh as the reference. The CD value is computed as follows:
\begin{equation}
\label{equation_CD}
\mathrm{CD}(P, P^{\prime}) = \frac{1}{|P|} \sum \limits_{\textbf{p}_i \in P} \min \limits_{\textbf{p}_j \in P^{\prime}} {||\textbf{p}_i - \textbf{p}_j||_{2}} 
+ \frac{1}{|P^{\prime}|} \sum \limits_{\textbf{p}_j \in P^{\prime}} \min \limits_{\textbf{p}_i \in P} {||\textbf{p}_j - \textbf{p}_i||_{2}},
\end{equation}
where $P$ and $P^{\prime}$ are densely sampled point clouds with $10^{5}$ points from the reconstructed mesh and the ground truth mesh, respectively. The normal consistency NC$(A,B)$ is calculated as the average of the absolute values of the dot products of the normals. Specifically, for each sample point in set $A$, we take its normal and compute the dot product with the normal of its nearest neighbor point in set $B$. The symmetric normal consistency is computed as $0.5 \times (\mathrm{NC}(A, B)+ \mathrm{NC}(B, A))$. The F-score is computed by evaluating the bidirectional correspondence between the reconstructed mesh and the ground truth mesh. Specifically, we first sample point clouds $P$ from the reconstruction and $P^{\prime}$ from the ground truth mesh. For each point $p_i$ in $P$, we compute its minimum Euclidean distance to any point in $P^{\prime}$. The precision is defined as the fraction of points in $P$ whose nearest-neighbor distance in $P^{\prime}$ is below a threshold $\tau = 7.5 \times 10^{-3}$ . Conversely, the recall is obtained by reversing the roles of $P$ and $P^{\prime}$. Finally, the F-score is derived as the harmonic mean of precision and recall as $(2 \times \mathrm{precision} \times \mathrm{recall}) / (\mathrm{precision} + \mathrm{recall})$. Each row in Table~\ref{table2} corresponds to the shapes shown in Figure~\ref{fig:Figure4b} and~\ref{fig:Figure4c}. In the last row of the table, we also present the average value calculated across the entire dataset, which consists of 486 shapes. The results clearly indicate the superior performance of our method.

In addition to traditional denoising methods, we also compare our method with Shape As Point (SAP)~\cite{2021DPSR} and GCNO~\cite{2023GCNO}. SAP proposes a differentiable Poisson solver for optimization and learning-based surface reconstruction. The main focus of SAP is different from our method. It only reconstructs surfaces with fixed point positions. In Figure~\ref{fig:Figure8}, we present a comparison between our method and SAP using two parameter settings for SAP: $\sigma=2.0$ (the default parameter) and $\sigma=7.0$. A large $\sigma$ smoothens the surface by filtering the Poisson field. However, it also introduces ripples that do not correspond to the surface texture, as can be seen in the teapot example. Our method optimizes point positions, leading to surfaces with noticeable denoising effects. Since the learning-based method of SAP involves a neural network and requires a large amount of training data, we only conduct comparisons with the optimization-based method.

GCNO~\cite{2023GCNO} is a normal orientation technique based on the winding number theory and requires screened Poisson surface reconstruction~\cite{2013SPSR} to generate a triangle mesh from the oriented point set. While both GCNO and our method aim to generate the expected winding number field, GCNO focuses on achieving this through normal orientation. Our approach goes further by evaluating how well a raw point cloud can form a good implicit field. We define the residual between the real field and the expected field as the winding clearness error, which serves both as a quality measure for the point cloud and as an optimization objective for point cloud denoising. In Figure~\ref{fig:Figure10}, we show the results of directly orienting and reconstructing a noisy point cloud using GCNO (denoted as ``GCNO+SPSR''), as well as the results obtained by denoising with our method before orientation and reconstruction with GCNO and SPSR (labeled as “Ours+GCNO+SPSR”). GCNO is executed with default parameters. The results demonstrate the effectiveness of our approach on thin structures and multiple connected components.

In Appendix B, we compare our method with other unoriented reconstruction approaches PGR~\cite{2022PGR} and WNNC~\cite{2024WNNC}, and outline the advantages of direct denoising over the ``reconstruction then project'' strategy. In Appendix D, we also compare our method with Moving Least Squares (MLS) surfaces and learning-based denoising techniques. Additionally, we conduct ablation studies on various parameters of our method and perform experiments on misaligned data, non-uniformly sampled data, data with missing regions, and real scanned data.

\subsection{Effectiveness of winding clearness constraint in diffusion-based generation}\label{sec:6_3}
In this section, we conduct experiments to evaluate the efficacy of the winding clearness constraint in diffusion-based generation. We adopt PVD as the baseline~\cite{2021PVD}, which entails training a shape generation model using standard DDPM on the ShapeNet dataset~\cite{2015ShapeNet}. Each generated point cloud comprises 2048 points. The results of the velocity-based diffusion can be found in Appendix C.2.

The original implementation of PVD offers several pre-trained models, like ``chair\_1799.pth'' and ``car\_3999.pth'', wherein the number represents the trained epoch minus one. We fine-tune these models using the joint training strategy proposed in Section~\ref{sec:52}. The learning rate is set to $2e^{-7}$, and $\rho$ in Eq.~\eqref{eq:gradient} is set to $1.0$. The experiments are carried out on a single NVIDIA GeForce RTX $4090$ GPU. We fine-tune the chair model for an additional $13$ epochs and the car model for $26$ epochs with the winding clearness constraint. Each training process takes approximately $26$ hours. 

In addition to the experiment of the single category, we test our method for generation across all $55$ classes. Given that the authors do not release the model for generating all $55$ classes, we initially trained the models for $400$ epochs with a decayed learning rate of 0.995 and keep all other parameters consistent with the source code. We then fine-tune the models for an additional $3$ epochs within $46$ hours using our method.

The quantitative comparisons are shown in Table~\ref{table3}, where we demonstrate the results for the chair, car category and all the 55 classes. The baseline results are labeled as ``PVD'' and the results obtained by our method is denoted as ``PVD+Ours''. To ensure a fair comparison, we conduct two types of ablation studies, labeled as “PVD+Ablation” and “PVD+Ablation*”, respectively. In “PVD+Ablation”, our method is executed with $\rho=0$ in Eq.~\eqref{eq:gradient} to exclude the contribution of the winding clearness term. Equal numbers of epochs are used for ablations ($13$ for chairs, $26$ for cars, and $3$ for 55 classes). 

Incorporating the winding clearness requires a reverse diffusion process, resulting in a longer period for an epoch when fine-tuning the model. Therefore, in “PVD+Ablation*”, we demonstrate that prolonged original training time without winding clearness does not necessarily enhance the quality of the generated point clouds. We present the generated point clouds for chairs, cars, and all $55$ classes at training epochs $3600$, $6600$, and $1000$ in “PVD+Ablation*”. The additional training time for these epochs exceeds the time required for the fine-tuning.

During testing, we employ the same initial Gaussian prior $\tilde{\textbf{x}}_0$ and intermediate Gaussian noise $\textbf{z}_t$ across all methods to generate the point clouds. The evaluation metrics include the winding clearness error (WCE) and the mean absolute distance-to-surface (MADS) between the point cloud and the generated surface of itself using PGR~\cite{2022PGR}. The WCE can be computed by Eq.~\eqref{eq:mu} and~\eqref{eq:wp}, while the MADS is computed as follows:
\begin{equation}
\label{equation_MADS}
\mathrm{MADS}(P, P^{\prime}) = \frac{1}{|P|} \sum \limits_{\textbf{p}_i \in P} \min \limits_{\textbf{p}_j \in P^{\prime}} {||\textbf{p}_i - \textbf{p}_j||_{2}},
\end{equation}
where $P$ is the generated point cloud and $P^{\prime}$ is a densely sampled point cloud with $10^{5}$ points from the reconstructed mesh of $P$. Table \ref{table3} shows a noticeable enhancement in the quality of generated point sets through ``PVD+Ours'', evidenced by lower WCE and MADS values. Meanwhile, the results obtained from ``PVD+Ablation'' and ``PVD+Ablation*'' do not exhibit considerable improvements compared to ``PVD''. The training objective of the diffusion model mainly focuses on converging to the distribution of the training set, rather than optimizing the quality of the point clouds. Consequently, simply extending the training period without incorporating the winding clearness does not lead to considerable improvement in point cloud quality.

\begin{table}[t]
\centering
\caption{Quantitative comparisons of PVD generation with and without the winding clearness constraint. The winding clearness error (WCE) and the mean absolute distance-to-surface (MADS) values are multiplied by $10^{3}$. }
\label{table3}
\resizebox{\linewidth}{!}{
\begin{tabular}{ccccccc}
\toprule
\multirow{2}{*}{Method} & \multicolumn{2}{c}{Chair} & \multicolumn{2}{c}{Car} & \multicolumn{2}{c}{55 Classes}\cr
\cmidrule(lr){2-7}

  & WCE $\downarrow$ & MADS $\downarrow$ & WCE $\downarrow$ & MADS $\downarrow$ & WCE $\downarrow$ & MADS $\downarrow$\\
\midrule

PVD & $4.35$ & $5.40$ &  $4.39$ & $7.33$ &  $4.64$ & $6.40$\\

PVD+Ablation & $4.42$ & $5.58$ &  $4.45$ & $7.45$ &  $4.70$ & $6.55$\\

PVD+Ablation* & $4.38$ & $5.46$ &  $4.38$ & $7.31$ &  $4.68$ & $6.38$\\

PVD+Ours & $\textbf{3.65}$ & $\textbf{4.28}$ & $\textbf{3.45}$ & $\textbf{6.18}$ & $\textbf{3.57}$ & $\textbf{4.65}$\\

\bottomrule

\end{tabular}
}
\end{table}

Figure~\ref{fig:Figure6} displays the qualitative results, where we show the shapes generated within the same initial Gaussian prior and intermediate noises. The only difference is the network parameters. We also colorize the distance from point clouds to the generated surfaces. The distance is noticeably smaller after the fine-tuning with the winding clearness constraint, which demonstrates the improvement in point cloud quality achieved by our method.

\section{Limitation and Future Work}
In this work, we propose winding clearness to represent the clarity of the interior/exterior relationship formed by the winding number field. We outline the formulation for calculating the winding clearness error and utilize it as a loss function for differentiable point cloud optimization. Our method significantly enhances the overall quality of the point cloud by optimizing the point positions to improve the winding clearness.

\begin{figure}[htb]
  \centering
  \includegraphics[width=1.0\linewidth]{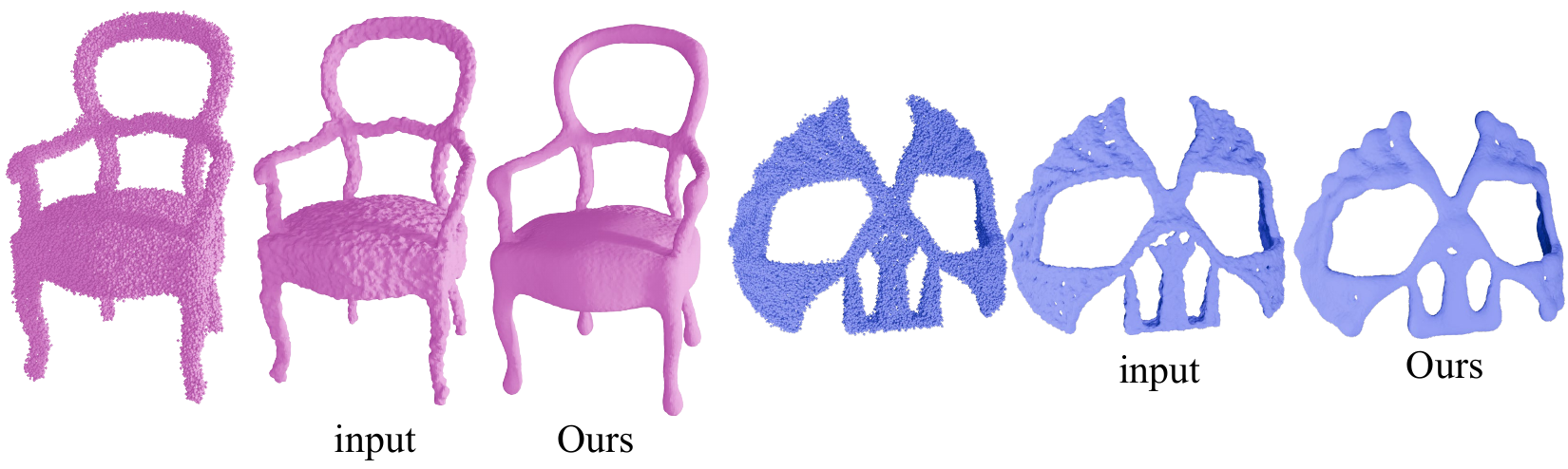}
  \caption{\label{fig50K}
           Results of our method on point clouds with $50$K points using multiple batch strategy.}
\end{figure}

Nevertheless, our method still has some limitations. The primary drawbacks are the time and memory requirements associated with our current implementation due to the global computation required by the winding number. Processing a point cloud with $5000$ points entails approximately $68$ s and utilizes $14$ GB of memory on the RTX 4090 GPU. The operation of $\textbf{A}^{\top} \textbf{A}$ has a time complexity of $O(N^{3})$ and a space complexity of $O(N^{2})$. However, we believe that we provide a fresh perspective for thin structure denoising with global awareness. One of the solutions for dealing with larger inputs is to process the point clouds in multiple batches. Specifically, we can randomly subsample 5000 points in the first batch and perform denoising based on the winding clearness. Then, we can randomly sample 2500 points from the undenoised point set and 2500 points from the denoised point set, and combine them to form matrices $\textbf{A}_1$ and $\textbf{A}_2$ in Eq.~\eqref{eq:fmu}. Next, the gradients are only propagated to the 2500 undenoised points. We repeat this process until the entire point cloud is fully denoised. Figure~\ref{fig50K} shows the denoising effect of the multiple batch strategy on point sets with $50$K points. We also use the reconstructed surface to demonstrate our denoising effects. A more advanced approach for large inputs could be to apply the FMM method~\cite{1987FMM, 2018FastWinding, 2019GR, 2024WNNC, 2024FastDipole}. Nevertheless, the movement of point positions during the optimization process must be taken into consideration.

Additionally, our method lacks a feature-preserving mechanism. A potential solution is to detect features before optimization~\cite{2024FeatureCurve} and assign higher weights to the feature points in Eq.~\eqref{eq:optimize}. In the future, we will continue to explore highly efficient point cloud processing with global awareness.

\section*{References}

\bibliography{mybibfile}

\begin{thebibliography}{10}
\expandafter\ifx\csname url\endcsname\relax
  \def\url#1{\texttt{#1}}\fi
\expandafter\ifx\csname urlprefix\endcsname\relax\def\urlprefix{URL }\fi
\expandafter\ifx\csname href\endcsname\relax
  \def\href#1#2{#2} \def\path#1{#1}\fi

\bibitem{2013Winding}
A.~Jacobson, L.~Kavan, O.~Sorkine{-}Hornung, Robust inside-outside segmentation using generalized winding numbers, {ACM} Trans. Graph. 32~(4) (2013) 33:1--33:12.

\bibitem{2022Medial}
N.~Wang, B.~Wang, W.~Wang, X.~Guo, Computing medial axis transform with feature preservation via restricted power diagram, {ACM} Trans. Graph. 41~(6) (2022) 188:1--188:18.

\bibitem{2022RDT}
P.~Wang, Z.~Wang, S.~Xin, X.~Gao, W.~Wang, C.~Tu, Restricted {D}elaunay triangulation for explicit surface reconstruction, {ACM} Trans. Graph. 41~(5) (2022) 180:1--180:20.

\bibitem{2023WindingDiscrete}
N.~Feng, M.~Gillespie, K.~Crane, Winding numbers on discrete surfaces, {ACM} Trans. Graph. 42~(4) (2023) 36:1--36:17.

\bibitem{2019GR}
W.~Lu, Z.~Shi, J.~Sun, B.~Wang, Surface reconstruction based on the modified {G}auss formula, {ACM} Trans. Graph. 38~(1) (2019) 2:1--2:18.

\bibitem{2022PGR}
S.~Lin, D.~Xiao, Z.~Shi, B.~Wang, Surface reconstruction from point clouds without normals by parametrizing the {G}auss formula, {ACM} Trans. Graph. 42~(2) (2023) 14:1--14:19.

\bibitem{2021dipole}
G.~Metzer, R.~Hanocka, D.~Zorin, R.~Giryes, D.~Panozzo, D.~Cohen{-}Or, Orienting point clouds with dipole propagation, {ACM} Trans. Graph. 40~(4) (2021) 165:1--165:14.

\bibitem{2023GCNO}
R.~Xu, Z.~Dou, N.~Wang, S.~Xin, S.~Chen, M.~Jiang, X.~Guo, W.~Wang, C.~Tu, Globally consistent normal orientation for point clouds by regularizing the winding-number field, {ACM} Trans. Graph. 42~(4) (2023) 111:1--111:15.

\bibitem{2024WNNC}
S.~Lin, Z.~Shi, Y.~Liu, Fast and globally consistent normal orientation based on the winding number normal consistency, {ACM} Trans. Graph. 43~(6) (2024) 189:1--189:19.

\bibitem{2017ReconSurvey}
M.~Berger, A.~Tagliasacchi, L.~M. Seversky, P.~Alliez, G.~Guennebaud, J.~A. Levine, A.~Sharf, C.~T. Silva, A survey of surface reconstruction from point clouds, Comput. Graph. Forum 36~(1) (2017) 301--329.

\bibitem{2021DPM}
S.~Luo, W.~Hu, Diffusion probabilistic models for 3d point cloud generation, in: {IEEE} Conference on Computer Vision and Pattern Recognition, Computer Vision Foundation / {IEEE}, 2021, pp. 2837--2845.

\bibitem{2021PVD}
L.~Zhou, Y.~Du, J.~Wu, 3{D} shape generation and completion through {P}oint-{V}oxel diffusion, in: 2021 {IEEE/CVF} International Conference on Computer Vision, {IEEE}, 2021, pp. 5806--5815.

\bibitem{2022LION}
X.~Zeng, A.~Vahdat, F.~Williams, Z.~Gojcic, O.~Litany, S.~Fidler, K.~Kreis, {LION}: Latent point diffusion models for 3d shape generation, in: Annual Conference on Neural Information Processing Systems, 2022.

\bibitem{2023FastDiffusion}
L.~Wu, D.~Wang, C.~Gong, X.~Liu, Y.~Xiong, R.~Ranjan, R.~Krishnamoorthi, V.~Chandra, Q.~Liu, Fast point cloud generation with straight flows, in: {IEEE/CVF} Conference on Computer Vision and Pattern Recognition, {IEEE}, 2023, pp. 9445--9454.

\bibitem{2018FastWinding}
G.~Barill, N.~G. Dickson, R.~M. Schmidt, D.~I.~W. Levin, A.~Jacobson, Fast winding numbers for soups and clouds, {ACM} Trans. Graph. 37~(4) (2018) 43.

\bibitem{1995PDEbook}
G.~B. Folland, Introduction to Partial Differential Equations, 2nd Edition, Princeton University Press, 1995.

\bibitem{2016Solid}
Q.~Zhou, E.~Grinspun, D.~Zorin, A.~Jacobson, Mesh arrangements for solid geometry, {ACM} Trans. Graph. 35~(4) (2016) 39:1--39:15.

\bibitem{2014FacetsOrientation}
K.~Takayama, A.~Jacobson, L.~Kavan, O.~Sorkine{-}Hornung, Consistently orienting facets in polygon meshes by minimizing the {D}irichlet energy of generalized winding numbers, CoRR abs/1406.5431.
\newblock \href {http://arxiv.org/abs/1406.5431} {\path{arXiv:1406.5431}}.

\bibitem{1987FMM}
L.~Greengard, V.~Rokhlin, A fast algorithm for particle simulations, Journal of Computational Physics 73~(2) (1987) 325--348.

\bibitem{2024FastDipole}
H.~Chen, B.~Miller, I.~Gkioulekas, 3{D} reconstruction with fast dipole sums, {ACM} Trans. Graph. 43~(6) (2024) 192:1--192:19.

\bibitem{2017ReviewFilter}
X.~Han, J.~S. Jin, M.~Wang, W.~Jiang, L.~Gao, L.~Xiao, A review of algorithms for filtering the 3{D} point cloud, Signal Process. Image Commun. 57 (2017) 103--112.

\bibitem{2022PointCloudReview}
L.~Zhou, G.~Sun, Y.~Li, W.~Li, Z.~Su, Point cloud denoising review: from classical to deep learning-based approaches, Graph. Model. 121 (2022) 101140.

\bibitem{2017Bilateral}
J.~Digne, C.~de~Franchis, The bilateral filter for point clouds, Image Process. Line 7 (2017) 278--287.

\bibitem{2019BilateralPCA}
F.~Zhang, C.~Zhang, H.~Yang, L.~Zhao, Point cloud denoising with principal component analysis and a novel bilateral filter, Traitement du Signal 36~(5) (2019) 393--398.

\bibitem{2005Jets}
F.~Cazals, M.~Pouget, Estimating differential quantities using polynomial fitting of osculating jets, Comput. Aided Geom. Des. 22 (2005) 121--146.

\bibitem{2007LOP}
Y.~Lipman, D.~Cohen{-}Or, D.~Levin, H.~Tal{-}Ezer, Parameterization-free projection for geometry reconstruction, {ACM} Trans. Graph. 26~(3) (2007) 22.

\bibitem{2009WLOP}
H.~Huang, D.~Li, H.~Zhang, U.~M. Ascher, D.~Cohen{-}Or, Consolidation of unorganized point clouds for surface reconstruction, {ACM} Trans. Graph. 28~(5) (2009) 176.

\bibitem{2003PointSetSurface}
M.~Alexa, J.~Behr, D.~Cohen{-}Or, S.~Fleishman, D.~Levin, C.~T. Silva, Computing and rendering point set surfaces, {IEEE} Trans. Vis. Comput. Graph. 9~(1) (2003) 3--15.

\bibitem{2005MovingSharp}
S.~Fleishman, D.~Cohen{-}Or, C.~T. Silva, Robust moving least-squares fitting with sharp features, {ACM} Trans. Graph. 24~(3) (2005) 544--552.

\bibitem{2009KernelRegre}
A.~C. {\"{O}}ztireli, G.~Guennebaud, M.~H. Gross, Feature preserving point set surfaces based on non-linear kernel regression, Comput. Graph. Forum 28~(2) (2009) 493--501.

\bibitem{2007APSS}
G.~Guennebaud, M.~H. Gross, Algebraic point set surfaces, {ACM} Trans. Graph. 26~(3) (2007) 23.

\bibitem{2010l1sparse}
H.~Avron, A.~Sharf, C.~Greif, D.~Cohen{-}Or, $l_{1}$-sparse reconstruction of sharp point set surfaces, {ACM} Trans. Graph. 29~(5) (2010) 135:1--135:12.

\bibitem{2015l0sparse}
Y.~Sun, S.~Schaefer, W.~Wang, Denoising point sets via ${L}_{0}$ minimization, Comput. Aided Geom. Des. 35-36 (2015) 2--15.

\bibitem{2022Elliptic}
A.~Agathos, P.~N. Azariadis, S.~Kyratzi, Elliptic {G}abriel {T}aubin smoothing of point clouds, Comput. Graph. 106 (2022) 20--32.

\bibitem{2022LowRank}
X.~Lu, S.~Schaefer, J.~Luo, L.~Ma, Y.~He, Low rank matrix approximation for 3{D} geometry filtering, {IEEE} Trans. Vis. Comput. Graph. 28~(4) (2022) 1835--1847.

\bibitem{2020Pointcleannet}
M.~Rakotosaona, V.~L. Barbera, P.~Guerrero, N.~J. Mitra, M.~Ovsjanikov, {P}oint{C}lean{N}et: Learning to denoise and remove outliers from dense point clouds, Comput. Graph. Forum 39~(1) (2020) 185--203.

\bibitem{2021Pointfilter}
D.~Zhang, X.~Lu, H.~Qin, Y.~He, Pointfilter: Point cloud filtering via encoder-decoder modeling, {IEEE} Trans. Vis. Comput. Graph. 27~(3) (2021) 2015--2027.

\bibitem{2022MODNet}
A.~Huang, Q.~Xie, Z.~Wang, D.~Lu, M.~Wei, J.~Wang, {MODN}et: Multi-offset point cloud denoising network customized for multi-scale patches, Comput. Graph. Forum 41~(7) (2022) 109--119.

\bibitem{2023PCDNF}
Z.~Liu, S.~Zhan, Y.~Zhao, Y.~Liu, R.~Chen, Y.~He, {PCDNF}: Revisiting learning-based point cloud denoising via joint normal filtering, CoRR abs/2209.00798.
\newblock \href {http://arxiv.org/abs/2209.00798} {\path{arXiv:2209.00798}}.

\bibitem{2023IterativePFN}
D.~de~Silva~Edirimuni, X.~Lu, Z.~Shao, G.~Li, A.~Robles{-}Kelly, Y.~He, Iterative{PFN}: True iterative point cloud filtering, in: {IEEE/CVF} Conference on Computer Vision and Pattern Recognition, {IEEE}, 2023, pp. 13530--13539.

\bibitem{2024SharpNeural}
C.~Chen, Y.~Liu, Z.~Han, Sharpening neural implicit functions with frequency consolidation priors, CoRR abs/2412.19720.
\newblock \href {http://arxiv.org/abs/2412.19720} {\path{arXiv:2412.19720}}.

\bibitem{2013EdgeAware}
H.~Huang, S.~Wu, M.~Gong, D.~Cohen{-}Or, U.~M. Ascher, H.~R. Zhang, Edge-aware point set resampling, {ACM} Trans. Graph. 32~(1) (2013) 9:1--9:12.

\bibitem{2022RFEPS}
R.~Xu, Z.~Wang, Z.~Dou, C.~Zong, S.~Xin, M.~Jiang, T.~Ju, C.~Tu, {RFEPS}: Reconstructing feature-line equipped polygonal surface, {ACM} Trans. Graph. 41~(6) (2022) 228:1--228:15.

\bibitem{2018MeshDenoising}
M.~Centin, A.~Signoroni, Mesh denoising with (geo)metric fidelity, {IEEE} Trans. Vis. Comput. Graph. 24~(8) (2018) 2380--2396.

\bibitem{2020DDPM}
J.~Ho, A.~Jain, P.~Abbeel, Denoising diffusion probabilistic models, in: Annual Conference on Neural Information Processing Systems, 2020.

\bibitem{2023DiffusionSDF}
G.~Chou, Y.~Bahat, F.~Heide, Diffusion-{SDF}: Conditional generative modeling of signed distance functions, in: {IEEE/CVF} International Conference on Computer Vision, {IEEE}, 2023, pp. 2262--2272.

\bibitem{2023ControllableDiffusion}
Z.~Lyu, J.~Wang, Y.~An, Y.~Zhang, D.~Lin, B.~Dai, Controllable mesh generation through sparse latent point diffusion models, in: {IEEE/CVF} Conference on Computer Vision and Pattern Recognition, {IEEE}, 2023, pp. 271--280.

\bibitem{2000Surfel}
H.~Pfister, M.~Zwicker, J.~van Baar, M.~Gross, Surfels: Surface elements as rendering primitives, SIGGRAPH '00, ACM Press/Addison-Wesley Publishing Co., USA, 2000, p. 335–342.

\bibitem{2003Strang}
B.~G. Strang, Introduction to linear algebra: Third edition, Introduction to Linear Algebra 28~(Supplement S1) (2003) S33–S47.

\bibitem{1952CG}
M.~R. Hestenes, E.~L. Stiefel, Methods of conjugate gradients for solving linear systems, Journal of Research of the National Bureau of Standards (United States) 49.

\bibitem{2019pytorch}
A.~Paszke, S.~Gross, et~al., Py{T}orch: an imperative style, high-performance deep learning library, CoRR abs/1912.01703.
\newblock \href {http://arxiv.org/abs/1912.01703} {\path{arXiv:1912.01703}}.

\bibitem{2014Adam}
D.~P. Kingma, J.~Ba, Adam: {A} method for stochastic optimization, in: Y.~Bengio, Y.~LeCun (Eds.), 3rd International Conference on Learning Representations, 2015.

\bibitem{2019PVCNN}
Z.~Liu, H.~Tang, Y.~Lin, S.~Han, Point-{V}oxel {CNN} for efficient 3d deep learning, in: Annual Conference on Neural Information Processing Systems, 2019, pp. 963--973.

\bibitem{2022Survey}
Z.~Huang, Y.~Wen, Z.~Wang, J.~Ren, K.~Jia, Surface reconstruction from point clouds: a survey and a benchmark, {IEEE} Trans. Pattern Anal. Mach. Intell. 46~(12) (2024) 9727--9748.

\bibitem{3DNet2012}
W.~Wohlkinger, A.~Aldoma, R.~B. Rusu, M.~Vincze, 3{DNet}: Large-scale object class recognition from {CAD} models, in: {IEEE} International Conference on Robotics and Automation, 2012, pp. 5384--5391.

\bibitem{2019ABC}
S.~Koch, A.~Matveev, Z.~Jiang, F.~Williams, A.~Artemov, E.~Burnaev, M.~Alexa, D.~Zorin, D.~Panozzo, A{BC}: a big {CAD} model dataset for geometric deep learning, in: Proceedings of the IEEE/CVF Conference on Computer Vision and Pattern Recognition (CVPR), 2019.

\bibitem{2016Thingi10K}
Q.~Zhou, A.~Jacobson, Thingi10{K}: {A} dataset of 10, 000 3d-printing models, CoRR abs/1605.04797.
\newblock \href {http://arxiv.org/abs/1605.04797} {\path{arXiv:1605.04797}}.

\bibitem{20213DScans}
{Oliver, L. et al.}, \href{https://threedscans.com}{Three {D} {S}cans: {F}ree 3{D} scan archive}, 2012.
\newline\urlprefix\url{https://threedscans.com}

\bibitem{2021DPSR}
S.~Peng, C.~Jiang, Y.~Liao, M.~Niemeyer, M.~Pollefeys, A.~Geiger, Shape {A}s {P}oints: a differentiable poisson solver, in: Annual Conference on Neural Information Processing Systems, 2021, pp. 13032--13044.

\bibitem{2009CGAL}
A.~Fabri, S.~Pion, {CGAL:} the computational geometry algorithms library, in: 17th {ACM} {SIGSPATIAL} International Symposium on Advances in Geographic Information Systems, {ACM}, 2009, pp. 538--539.

\bibitem{2022iPSR}
F.~Hou, C.~Wang, W.~Wang, H.~Qin, C.~Qian, Y.~He, Iterative {P}oisson surface reconstruction (i{PSR}) for unoriented points, {ACM} Trans. Graph. 41~(4) (2022) 128:1--128:13.

\bibitem{2013SPSR}
M.~M. Kazhdan, H.~Hoppe, Screened {P}oisson surface reconstruction, {ACM} Trans. Graph. 32~(3) (2013) 29:1--29:13.

\bibitem{2015ShapeNet}
A.~X. Chang, T.~A. Funkhouser, et~al., Shape{N}et: an information-rich 3{D} model repository, CoRR abs/1512.03012.
\newblock \href {http://arxiv.org/abs/1512.03012} {\path{arXiv:1512.03012}}.

\bibitem{2024FeatureCurve}
U.~Fugacci, C.~Romanengo, B.~Falcidieno, S.~Biasotti, Reconstruction and preservation of feature curves in 3{D} point cloud processing, Comput. Aided Des. 167 (2024) 103649.

\end{thebibliography}
\clearpage
\includepdf[pages=-]{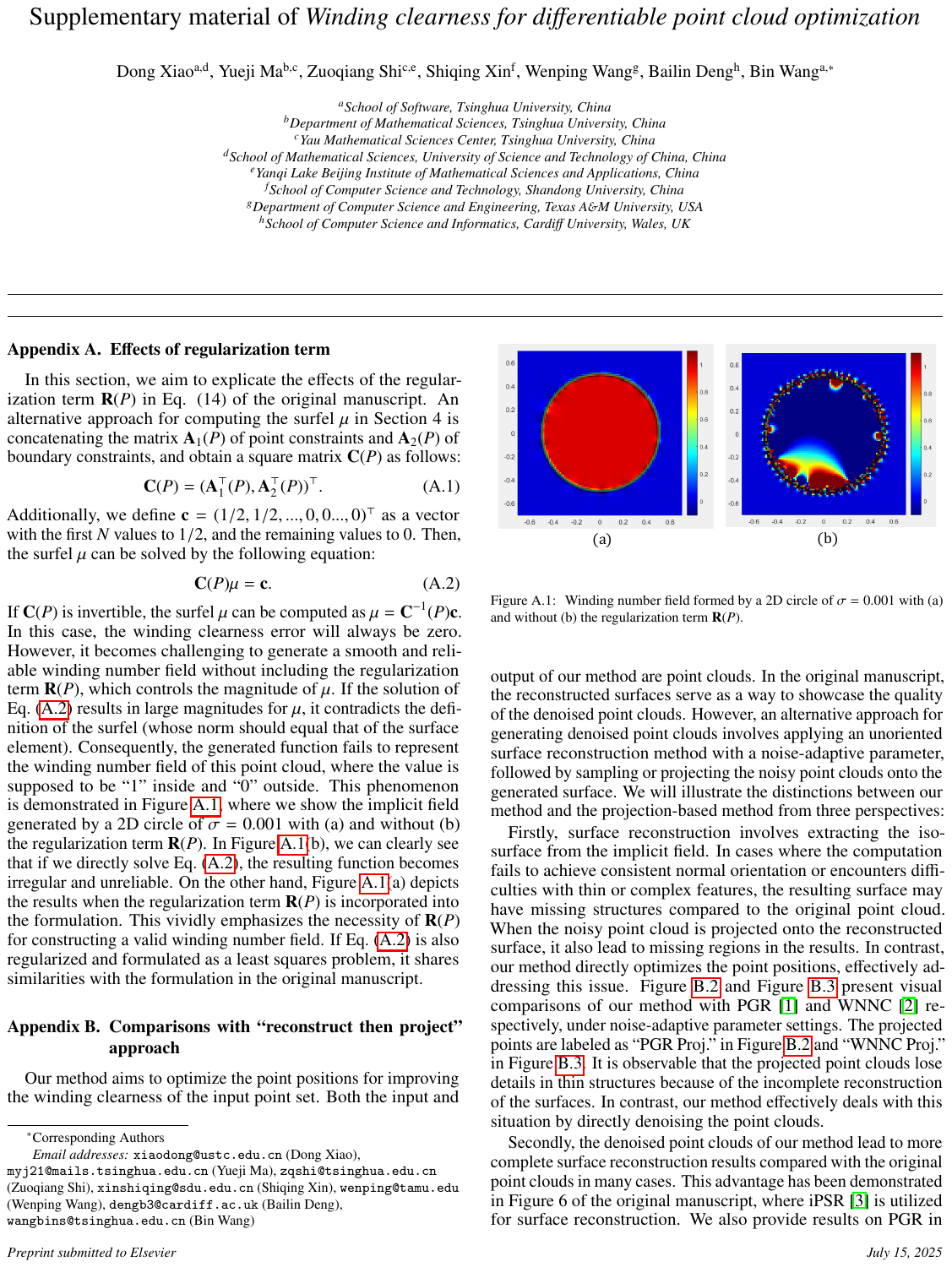}
\end{document}